%% file: main.tex
\pdfoutput=1
\documentclass[a4paper, amsfonts, amssymb, amsmath, reprint, showkeys, nofootinbib, twoside, superscriptaddress]{revtex4-1}
\usepackage{comment}
\usepackage{graphicx}
\usepackage{dcolumn}
\usepackage{bm}
\usepackage{amsfonts}
\usepackage{amssymb}
\usepackage[acronym]{glossaries}
\usepackage{rotating}
\usepackage{xcolor}

\usepackage{dutchcal}
\DeclareMathAlphabet{\mathdutchcal}{U}{dutchcal}{m}{n}
\SetMathAlphabet{\mathdutchcal}{bold}{U}{dutchcal}{b}{n}
\DeclareMathAlphabet{\mathdutchbcal}{U}{dutchcal}{b}{n}

\usepackage{multirow}
\usepackage{amsthm}

\usepackage{thmtools}
\usepackage{thm-restate}

\usepackage{hyperref}
\hypersetup{colorlinks,allcolors=blue}

\newcommand{\ket}[1]{|#1\rangle}

\def\be{\begin{equation}} %
\def\ee{\end{equation}} %
\newcommand{\bea}{\begin{eqnarray}}
\newcommand{\eea}{\end{eqnarray}}

\newcommand{\blue}[1]{\textcolor{black}{#1}}

\begin{document}
\title{Block-Invariant Symmetry Shift: Preprocessing technique for second-quantized Hamiltonians to improve their decompositions to Linear Combination of Unitaries}

\author{Ignacio Loaiza}
\affiliation{University of Toronto, Department of Chemistry, Chemical Physics Theory Group, Toronto, Canada}
\affiliation{University of Toronto Scarborough, Department of Physical and Environmental Sciences, Toronto, Canada}
\affiliation{Zapata Computing Canada Inc., Toronto, Canada}
\author{Artur F. Izmaylov}
\email{artur.izmaylov@utoronto.ca}
\affiliation{University of Toronto, Department of Chemistry, Chemical Physics Theory Group, Toronto, Canada}
\affiliation{University of Toronto Scarborough, Department of Physical and Environmental Sciences, Toronto, Canada}
\date{\today}

\newacronym{CSA}{CSA}{Cartan sub-algebra}
\newacronym{LCU}{LCU}{linear combination of unitaries}
\newacronym{BLISS}{BLISS}{block-invariant symmetry shift}
\newacronym{QPE}{QPE}{quantum phase estimation}
\newacronym{DF}{DF}{double factorization}

\begin{abstract}
Computational cost of energy estimation for molecular electronic Hamiltonians via \gls{QPE} grows with the \blue{difference between the largest and smallest eigenvalues} of the Hamiltonian. In this work we propose a preprocessing procedure that reduces the 
norm of the Hamiltonian without changing its eigenspectrum for the target states of a particular symmetry. 
The new procedure, \gls{BLISS}, builds an operator $\hat T$ such that the cost of implementing 
$\hat H - \hat T$ is reduced compared to that of $\hat{H}$, yet $\hat H - \hat T$ acts on the subspaces of interest the same way as $\hat H$ does. \gls{BLISS} performance is demonstrated for 
\gls{LCU}-based \gls{QPE} approaches on a set of small molecules. 
Using the number of electrons as the symmetry specifying the target set of states, \gls{BLISS} provided a factor of \blue{2 reduction of 1-norm for several \gls{LCU} decompositions compared to their unshifted versions.}
\end{abstract}

\maketitle

\glsresetall

\section{Introduction}
Quantum chemistry is a promising field where quantum computers can potentially solve useful problems that are challenging for classical computers. One of the distinctive advantages of quantum computers lies in their linear qubit requirement for representing electronic degrees of freedom, in contrast to the exponential growth of classical bits that would be needed \cite{QCinQC, polynomial_dynamics}. This inherent scalability offers a potential solution to the challenges associated with managing high-dimensional systems. However, the focus shifts to efficiently preparing electronic wavefunctions of interest, such as the electronic ground state represented by the electronic structure Hamiltonian $\hat H$, as a key challenge for quantum computers in this domain.

\blue{Different versions of the \gls{QPE} method offer an efficient approach for performing energy estimations on fault-tolerant quantum computers. These techniques stand out due to their optimal scalability and guaranteed accuracy \cite{QPE,kitaev_QPE,aspuru_qpe,heisenberg_qpe,statistical_qpe,ts_qpe,cirac_qpe,GSEE}.}  
\blue{Phase estimation requires the propagator of a simple function of the Hamiltonian to be represented on a quantum circuit}, which corresponds to implementing a function $f(\hat H)$ specific to the quantum algorithm. Currently, two commonly used choices for $f(\hat H)$ are: 
1) $f(\hat H) = e^{-i\hat H t}$ and 2) $f(\hat H) = e^{i \arccos (\hat H/\alpha)}$, for $\alpha$ a normalizing constant. The implementation of the first function is done using either \gls{LCU}-based approaches \cite{LCU,interaction,qubitized_pe,THC} or Trotterization \cite{trotter}. The second function appears as a result of block-encoding of $\hat H$ as a part of a unitary operator in a larger qubit space, this constitutes so-called qubitization process\cite{low2019hamiltonian, qubitization_wiebe}.
The run times and circuit depths of these algorithms differ depending on $f(\hat H)$ implementations.

Run time estimates for \blue{the qubitization-based phase estimation} algorithm applied to some industrially relevant molecules (e.g. Li-ion battery compounds)\cite{xanadu_batteries} have highlighted the need for further improvements to make it practically applicable in quantum chemistry contexts \cite{xanadu_batteries, bloch_thc, mol_obs, ft_li_battery, p450}. Reducing the computational and resource costs of quantum algorithms can be achieved by narrowing the \blue{spectral range}, while keeping its spectrum unchanged for the eigenstates of interest. \blue{The spectral range is defined as $\Delta E \equiv E_{\max} - E_{\min}$, for $E_{\max(\min)}$ the maximum(minimum) eigenvalue of $\hat H$}. In this study, we consider the cost reduction aspects of QPE algorithms involving LCU. There are at least two components within these algorithms that can benefit from this norm reduction strategy.

First, \gls{QPE} algorithms exhibit a cost that scales with the target accuracy $\epsilon$. Approaches with optimal scaling achieve Heisenberg scaling $\tilde{\mathcal{O}}(\epsilon^{-1})$ \cite{heisenberg_qpe}. However, all these methods require a rescaling of the Hamiltonian to confine its spectrum within a fixed range, e.g. $[0,1]$, avoiding an aliasing for the recovered energy. This entails working with a scaled Hamiltonian
\begin{equation}
    \hat H_{sc} = \frac{\hat H - E_{\min}\hat 1}{\Delta E}.
\end{equation}
The phase estimation procedure guarantees an accuracy $\epsilon_{sc} = \epsilon/\Delta E$ for this scaled Hamiltonian. Thus, if a target accuracy $\epsilon$ is required for the energy, the associated accuracy for the phase estimation procedure becomes $\epsilon \Delta E$. This shows how reducing the spectral range of $\hat H$ will lower the cost of the phase estimation procedure. Some approaches exist that are able to go beyond the Heisenberg scaling \cite{GSEE}. However, these still require an initial run of the Heisenberg-limited phase estimation, and will also benefit from a reduction of $\Delta E$. \blue{It is important to note that for larger molecules, determining the spectral range is a task of comparable difficulty to finding the ground state energy itself. For \gls{LCU}-based methods, the 1-norm $\lambda$ of the \gls{LCU} can serve as an upper bound for $\Delta E$ \cite{loaiza_lcu}. In practice, this results in the reported $\mathcal{O}(\lambda/\epsilon)$ scaling in Refs.~\citenum{THC,qubitized_pe}. However, this 1-norm might not be well-defined for Trotter-based methods, for which the phase estimation procedure will also benefit from a spectral range reduction. Calculating the appropriate rescaling factor for Trotterized Hamiltonians lies beyond the scope of this study. Regardless of the method used to encode the Hamiltonian, the cost of phase estimation will inevitably scale with the spectral range.}

Second, the cost of $\hat{H}$ block-encoding via \gls{LCU}, 
\begin{equation} \label{eq:lcu}
    \hat H = \sum_k c_k \hat U_k
\end{equation}
with $\hat U_k$ unitaries, is the 1-norm of the $c_k$ vector, $\lambda \equiv \sum_k |c_k|$. 
This 1-norm is the key metric for evaluating \gls{LCU} decompositions. The cost of implementing \gls{LCU}-based quantum algorithms scales linearly with this norm, up to polylogarithmic factors, depending on the particular $f(\hat H)$ \cite{LCU,interaction,loaiza_lcu}. It has been established \cite{loaiza_lcu} that the lower bound of the LCU 1-norm 
is a half of the Hamiltonian spectral range, $\lambda \geq \Delta E/2$, 
regardless of the chosen $\hat U_k$ operators in the LCU.
However, the full cost of implementing $f(\hat H)$ also depends on other factors, such as the type of unitaries in the \gls{LCU}, their compilation strategy, \blue{and the practical implementation of the corresponding quantum circuit on actual quantum hardware} \cite{LCU, THC, femoco_df, qubitization_wiebe}.

In this work, we expand upon our previously proposed symmetry shift technique \cite{loaiza_lcu} and illustrate its dual advantage in diminishing both the quantum algorithm cost through $\Delta E$ reduction and the Hamiltonian encoding expense for \gls{LCU}-based methods. 

The central concept behind the symmetry shift method involves substituting the Hamiltonian with a symmetry-shifted counterpart, denoted as $\hat H - \hat S$, such that its action on subspaces of interest, i.e. wavefunctions with a target number of electrons, remains invariant. The construction of $\hat S$ entails minimizing the implementation cost of $\hat H - \hat S$. Here, we introduce the \gls{BLISS} technique, which subtracts an operator $\hat T$ that is not necessarily a symmetry 
of $\hat H$. This generates a new shifted Hamiltonian $\hat H_T \equiv \hat H - \hat T$ that still has an invariant action on subspaces of interest, meaning any algorithm can be run for $\hat H_T$ instead of $\hat H$, giving the same result with a lower implementation cost. Note that shifting the Hamiltonian with a symmetry has also been used to reduce costs for the Hubbard model \cite{campbell_hubbard}. However, our extension did not give any additional benefits for the Hubbard Hamiltonian, \blue{which we attribute to the sparse structure of this system: the majority terms affected by the symmetry shift were already zero.}\footnote{\blue{The Hubbard Hamiltonian is defined as
\begin{equation}
  \hat H = -J\sum_{\langle i,j \rangle}\sum_{\sigma=\alpha,\beta} (\hat a^\dagger_{i,\sigma} \hat a_{j,\sigma} + h.c.) + U\sum_{i=1}^N \hat n_{i\alpha}\hat n_{i\beta},
\end{equation}
where $\langle i,j\rangle$ indicates the sum is over nearest neighbours. Symmetries of this Hamiltonian that can be written as one- and two-electron operators correspond to the standard $\hat S_z, \hat S_+, \hat S_-, \hat S^2$, and $\hat N_e$ operators. Note how these act homogeneously over all orbital sites (e.g. $\hat S_z = \sum_i (\hat n_{i\alpha} - \hat n_{i\beta})/2$). Thus, any function of these operators that is able to modify components of the form $\hat n_{i\alpha}\hat n_{i\beta}$ will also affect already null components, e.g. $\hat n_{i\alpha}\sum_{j\neq i} \hat n_{j\beta}$. Whatever cost reduction was obtained by reducing the existing terms is hindered by the introduction of a large number of additional operators. The only cost reduction for the Hubbard Hamiltonian thus comes from the one-electron component, as considered in Ref.~\citenum{campbell_hubbard}.}} As a result, our focus is directed towards molecular electronic structure Hamiltonians. 

\section{Theory}
A spectral range reduction is only possible if we select a particular set of states whose eigenvalues 
will be invariant with respect to the 
modification and eigenvalues of other states will be allowed to change. 

Finding a transformation that achieves this requires using symmetry operators: they are the only operators besides the Hamiltonian whose action is simple on the Hamiltonian eigenstates. 
We consider a set of Hamiltonian eigenstates of interest $\{\ket{\psi_k}\}$, 
which means we know eigenvalues of symmetry 
operators for these states: $\hat S_i\ket{\psi_k} = s_i \ket{\psi_k} $. We can then build a modified Hamiltonian 
as 
\bea
\hat H_{m} = \hat{H} - f_m(\{\hat S_i-s_i\hat 1\}).
\eea
Independent of the shift function $f_m$, $\hat H_m \ket{\psi_k} = \hat H \ket{\psi_k} = E_k \ket{\psi_k}$. 
Now, we only need to select the shift $f_m$ so that the spectral range ($\Delta E$) and LCU 1-norms for $\hat H_m$  will be lower 
than those of $\hat H$. To achieve this we will use a simple heuristic measure, the LCU 1-norm of decomposing 
$\hat H_m$ as a linear combination of Pauli products. This measure is simple to evaluate and 
it correlates well with the quantities of interest, as found empirically from previous studies \cite{loaiza_lcu}. 
To carry out the optimization process to find a function $f_m$ we put several constraints: 1) $\hat H_m$ should be hermitian, 
2) expansions of $\hat H$ and $f_m(\{\hat S_i-s_i\hat 1\})$ in terms of fermionic operators have the same polynomial degrees, 
3) spin symmetry properties of electron-integral coefficients in $\hat H$ should not be altered by the modification. All these constraints are motivated by convenience of use and optimization of $\hat H_m$.

The electronic structure Hamiltonian can be written as
\begin{equation} \label{eq:mol_ham}
    \hat H = \sum_{ij}^N h_{ij} \hat F^i_j + \sum_{ijkl}^N g_{ijkl} \hat F^i_j \hat F^k_l,
\end{equation}
where $\{i,j,k,l\}$ are spacial orbitals, $h_{ij}$ and $g_{ijkl}$ are one- and two-electron integrals \footnote{Representing the Hamiltonian using only excitation operators is usually referred to as chemists' notation. This entails a modification to the one-electron tensor with respect to physicists' notation, which uses normal-ordered operators of the form $\hat a^\dagger_p\hat a^\dagger_q \hat a_r \hat a_s$. Our notation is related to the electronic integrals by $g_{ijkl} = \frac{1}{2}\int \int d\vec r_1 d\vec r_2 \frac{\phi_i^*(\vec r_1)\phi_j(\vec r_1)\phi_k(\vec r_2)\phi_l^*(\vec r_2)}{|\vec r_1 - \vec r_2|}$ and $h_{ij} =- \sum_k g_{ikkj} + \int d\vec r \phi_i^*(\vec r) \Big(-\frac{\nabla^2}{2} - \sum_n \frac{Z_n}{|\vec r - \vec R_n|} \Big) \phi_j(\vec r)$, with $\phi_i(\vec r)$ the one-particle electronic basis functions, and $Z_n/\vec R_n$ the charge/position of nucleus $n$.}, $N$ is the number of spacial one-electron orbitals, and $\hat F^i_j \equiv \sum_{\sigma} \hat a^\dagger_{i\sigma}\hat a_{j\sigma}$ are spacial excitation operators, with $\sigma\in\{\alpha,\beta\}$ $z$-spin projections. We note that writing the Hamiltonian with $\hat F^i_j$ operators allows us to work in spacial instead of spin-orbitals, greatly reducing the classical storage and manipulation cost of the corresponding fermionic tensors $h_{ij}$ and $g_{ijkl}$. We will refer to any operator that can be written in terms of $\hat F^i_j$'s as spin-symmetric. In addition, many existing \gls{LCU} methodologies use this spin-symmetric structure for both the \gls{LCU} decomposition and efficient compilation of the unitaries on a quantum computer \cite{THC,femoco_df}. Working with this spin-symmetric structure thus allows for existing approaches to be applied without any additional considerations.

In Ref.~\citenum{loaiza_lcu}, the choice for the shift was limited to  
\begin{equation}
    \hat S(\vec \kappa) \equiv \kappa_1(\hat N_e - N_e \hat 1) + \kappa_2(\hat N_e^2 - N_e^2 \hat 1),
\end{equation}
where $\hat N_e$ is the number of electron operator, $N_e$ is the target number of electrons, and $\kappa_u$'s are real parameters to be optimized. This operator corresponds to choosing the one- and two-electron symmetries $\hat N_e$ and $\hat N_e^2$ for $f_m$. One could also add $\hat S^2$, $\hat S_z$, $\hat S_z \hat N_e$, and $\hat S_z^2$ 
symmetry operators as one- and two-electron components, but they violate the 3$^{\rm rd}$ condition and do not 
provide a significant improvement. Thus, this symmetry-only shift gave  
\begin{equation} \label{eq:HS}
    \hat H_S(\vec\kappa) = \hat H - \hat S(\vec \kappa).
\end{equation}

Yet, one can extend the $f_m$ function to have operators that are not symmetries of the Hamiltonian, the simple idea of the 
\gls{BLISS} approach is to use products $\hat O_i (\hat S_i-s_i\hat 1)$ in $f_m$, where $\hat O_i$ are arbitrary hermitian operators  commuting with $\hat S_i$ so that the products are hermitian as well. Taking into account two-electron requirement for the shift operator the \gls{BLISS} approach provides 
\bea \label{eq:T}
    \hat T(\vec \kappa, \vec \xi) &\equiv& \hat S(\vec \kappa) + \sum_{ij} \xi_{ij} \hat F^i_j (\hat N_e - N_e\hat 1),\\
    \hat H_T(\vec \kappa,\vec \xi) &=& \hat H - \hat T(\vec \kappa, \vec \xi) \label{eq:bliss}
\eea
where $\vec \xi \equiv \{\xi_{ij}\}$ is a real vector with symmetric indices ($\xi_{ij}=\xi_{ji}$). 

It is worth mentioning that the elucidated \gls{BLISS} approach can be readily expanded to encompass molecular point-group symmetries, since these symmetries adhere to spin-symmetric one-electron operators \cite{molecular_point_group_symmetries}. However, when dealing with intricate molecules and transition-state geometries that tend to be the focus of quantum computing applications, point-group symmetries will often collapse to the identity operator. Therefore, we have not taken them into account within our considerations.

\blue{In the case where the initial state has significant overlap (e.g. greater than $0.5$) with the ground state, the \gls{BLISS} procedure can be used regardless of the symmetry of the initial state. To better understand this point, we consider the spectral distribution that is obtained as the output of a phase estimation procedure. Regardless of where the lowest energy peak is for $\hat H_T$, the ground state energy of the original Hamiltonian $\hat H$ will correspond to the highest peak in the distribution. Having an overlap that is greater than $0.5$ thus guaranties that there are no other peaks with a greater amplitude, although in practice we only require for the overlap with other eigenstates to be smaller than that with the ground state. This argument extends to the cumulative distribution function obtained in the Heisenberg-scaling phase estimation algorithm \cite{heisenberg_qpe}. Alternatively, there exist wavefunction preparation algorithms that do not break the number of electrons symmetry, namely: the unitary coupled cluster ansatz for the Variational Quantum Eigensolver (VQE)\cite{vqe,vqe_rev1,vqe_rev2,ucc}, VQE algorithms with a symmetry-breaking penalty term \cite{qcc}, or matrix product state preparation techniques \cite{mps_on_qc} using a classical number conserving ansatz such as Density Matrix Renormalization Group (DMRG) \cite{dmrg,dmrg_rev1,dmrg_rev2} or configuration interaction wavenfunctions \cite{ci_as_mps}. For wavefunctions that are eigenfunctions of the electron number operator, the lowest energy in the phase estimation procedure will correspond to the ground state energy of $\hat H$. Thus, these wavefunctions can be used even if the overlap is smaller than $0.5$. However, it is important to note that the full cost of the phase estimation procedure will increase as the quality of the initial wavefunction diminishes.}

\section{Results and discussion} \label{sec:discussion}
Table~\ref{tab:results} shows the spectral ranges and 1-norms for some small molecules and the different \gls{LCU} decompositions presented in the Appendix. For illustrating \gls{BLISS} performance we have used the number of electrons $N_e$ corresponding to a neutral molecular form, but any other charged state can be targeted as well. \blue{In order to succinctly show the 1-norm improvements from using \gls{BLISS}, a linear fit of the 1-norm changes over different molecules is shown in the last row of the table; the small reported standard errors justify the usage of a linear fit}. The linear fit was obtained by associating to each molecule two coordinates: $x$ is the 1-norm of the Pauli product LCU for the unshifted Hamiltonian, and $y$ is the 1-norm of the considered LCU method and Hamiltonian. The slope of the linear regression is given with the associated standard error ($\pm$), which corresponds to finding the best $\alpha$ such that $y \approx \alpha x$. \blue{In the interest of knowing the best possible 1-norm after \gls{BLISS} application, we included the symmetry-projected spectral range $\Delta E_{N_e}$ which corresponds to the spectral range of the Hamiltonian projected in the space with $N_e$ electrons.} Highlights show methods with best scaling. For $\Delta E/2$ \blue{and $\Delta E_{N_e}/2$}, the linear fit column corresponds to using the unshifted $\Delta E/2$ as the $x$-axis. Once the coefficient associated with the linear regression was found, a percentage of improvement was obtained as $(1-\alpha)\times 100\%$. In addition, we also discuss improvements of certain methods with respect to their unshifted versions. These were obtained by using the unshifted 1-norm of each method as the $x$-axis for the linear regression procedure, and correspond to the net cost reduction that is obtained by using the symmetry shifts with a given LCU decomposition technique.

\input{tables}
\begin{figure}
    \centering
    \includegraphics[width=8cm]{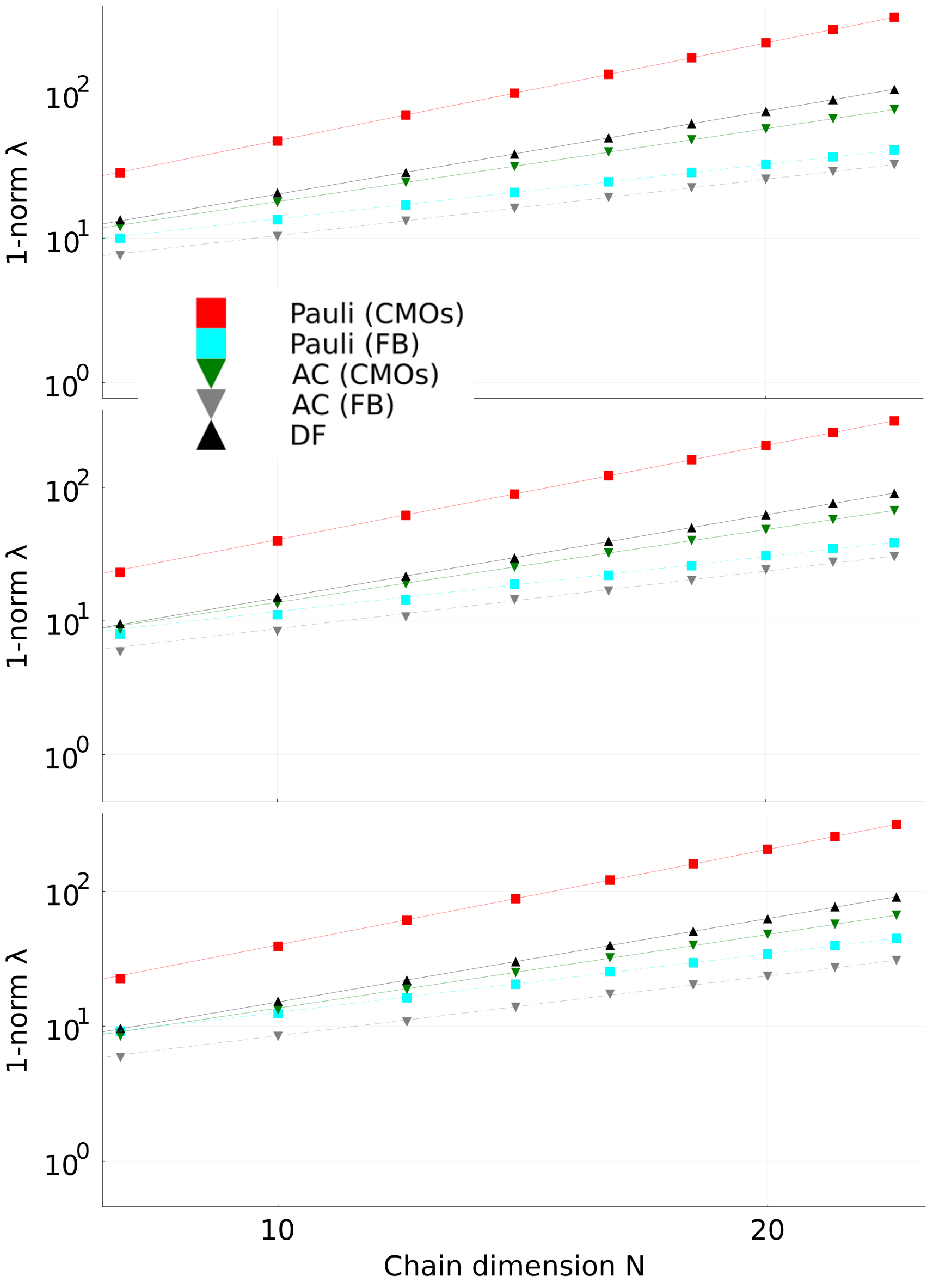}
    \caption{\blue{1-norms scaling in hydrogen chains with $N$ atoms and spacing $r=1.4$ \r{A} for various \gls{LCU}s and different Hamiltonians (top: $\hat H$, middle: $\hat H_S$, bottom: $\hat H_T$). Lines correspond to polynomial scaling fit, where we consider $\log_{10}\lambda = \alpha \log_{10} N + \beta$. Fit coefficients are reported in Table~\ref{tab:scaling}. CMOs(FB) correspond to Hartree-Fock canonical molecular (Foster-Boys localized) orbitals.}}
    \label{fig:scaling}
\end{figure}

For minimizing 1-norm of the Pauli product LCU of $\hat H_T(\vec\kappa,\vec \xi)$, we use a non-linear optimization package 
for the parameters $\vec \kappa$ and $\vec \xi$ \cite{optim}. 
Computational details of molecular electronic Hamiltonians and computational thresholds can be found in the Appendix. The orbital rotations for the orbital optimization scheme were obtained with a Broyden-Fletcher-Goldfarb-Shanno (BFGS) scheme for the Pauli LCU 1-norm in Table \ref{tab:results}. \blue{For orbitals in the hydrogen chains we have used the Foster-Boys localization scheme \cite{FB}. The Foster-Boys scheme is computationally more efficient than the orbital optimization that lowers 1-norm, while yielding 1-norm reductions that closely match the latter's outcomes \cite{majorana_l1}}. The anticommuting groups were obtained using a sorted insertion algorithm \cite{SI,anticommuting}, and the \gls{DF} fragments are found through a Cholesky decomposition of the two-electron tensor $g_{ijkl}$ \cite{df_1,df_2,df_3,df_4,df_5,THC}. The \gls{CSA} decomposition was done by obtaining fragments through a greedy optimization of their parameters one fragment at a time. \blue{The \gls{CSA} decomposition was not calculated for the hydrogen chains because it quickly becomes prohibitively expensive as the system size grows.}

Values in parenthesis in Table~\ref{tab:results} represent the total number of unitaries in the \gls{LCU}. Comparing the implementation costs of different LCUs is not a straightforward task, as the number and type of unitaries, in addition to the 1-norm, are linked to the implementation cost. Moreover, the specific form of the LCU decomposition can sometimes be used to construct more efficient circuits \cite{THC,femoco_df}, necessitating the explicit construction of the oracle circuits for a comprehensive comparison of different LCUs. This is particularly noticeable for the \gls{DF} method, with LCUs that have significantly less unitaries than the other decompositions but each unitary requires an implementation via qubitization. \blue{However, a full cost comparison of the quantum circuits is not necessary for showing the improvement from using \gls{BLISS}. Within \gls{LCU}-based encodings, the application of \gls{BLISS} yielded reductions in 1-norm values while maintaining the number of unitaries nearly intact: following the \gls{BLISS} procedure, there was an average decrease of approximately $2.2\%$ in the number of unitaries, with the maximum increase observed at $15.1\%$ for the NH$_3$ molecule with the orbital-optimized anticommuting grouping. Since the shifted Hamiltonian $\hat H_T$ has exactly the same operator form as $\hat H$, $\hat H$ and $\hat H_T$ can use the same compilation strategy. A similar number of unitaries in the \gls{LCU} decompositions of $\hat H$ and $\hat H_T$ means that the only cost difference in their implementations will arise from the change of 1-norm.}

We will refer to the previously developed shift technique \cite{loaiza_lcu} as a partial shift, in contrast to a full shift that corresponds to the \gls{BLISS} methodology and use of $\hat H_T$. We now start discussing the improvement of the spectral range $\Delta E$ when the \gls{BLISS} is applied, having an average improvement of $31\%$ and $42\%$ with respect to the initial spectral range for the partial and full shifts respectively. This shows how the application of the \gls{BLISS} will reduce significantly the quantum algorithm cost, requiring slightly more than half the resources of working with the full Hamiltonian $\hat H$. 

For the Hamiltonian encoding cost, all presented methods show a very significant improvement when using the partial shift, while having an additional improvement when the full \gls{BLISS} is used. Out of all studied methods, those with the best 1-norms are the orbital-optimized anticommuting grouping, and \gls{DF}, presenting an average improvement of $49\%$ when compared to their unshifted versions, while the partial shift gave an average improvement of  $38\%$ for \gls{DF} and of $40\%$ for the orbital-optimized anticommuting grouping. As such, the \gls{BLISS} technique practically halves the cost of implementing these \gls{LCU}-based encodings. 

From these results, it becomes clear that all of the previously mentioned energy estimation methods will greatly benefit from the application of the \gls{BLISS}. If a Trotter-based method is used for implementing $f(\hat H) = e^{-i\hat H t}$, the circuit cost and total run time will be reduced by $\sim42\%$ for algorithms that re-scale the Hamiltonian \cite{heisenberg_qpe}. However, if $f(\hat H)$ is to be implemented with an \gls{LCU}-based method, the block-encoding \gls{LCU}-based procedure encodes $f(\hat H/\lambda)$, which already has a normalized spectrum given how the 1-norm $\lambda \geq \Delta E/2$. Thus, the cost improvement from the encoding already incorporates the quantum algorithm cost reduction associated with re-scaling of $\hat H$. This results in a $\sim 49\%$ cost reduction for any \gls{LCU}-based algorithms, while the partial shift yields an improvement of $38-40\%$ for the qubit and fermionic-based techniques presented in this work. 

\blue{Furthermore, a comparison between the spectral range of the \gls{BLISS} modified Hamiltonian and that of the original Hamiltonian in the subspace with $N_e$ electrons reveals remarkably similar values. This observation shows that the shift outlined in Eq.\eqref{eq:bliss} cannot undergo significant further improvements: the \gls{BLISS} procedure effectively eliminates contributions originating from subspaces with different number of electrons. Figure \ref{fig:scaling} and Table \ref{tab:scaling} illustrate the 1-norm scaling pattern for an extended chain of hydrogen atoms. The scaling of the 1-norms as a function of the chain length shown in Table \ref{tab:scaling} changes insignificantly after application of the shift, which indicates that the 1-norm improvement will likely exhibit similar behavior for large systems. While generalizing these trends to larger molecules with more flexible one-particle basis sets might be intricate, the substantial reduction in phase estimation algorithm costs resulting from the elimination of contributions associated with varying electron counts remains evident. The \gls{BLISS} technique can be perceived as transitioning from the Hamiltonian in the complete Fock space to its counterpart within a fixed number of electrons subspace while remaining in a second quantization formalism. The question of whether first quantized Hamiltonians can similarly gain from symmetry shifts remains a topic under ongoing exploration.}

We note that combining the \gls{BLISS} technique with the previously proposed interaction picture methodology \cite{loaiza_lcu, interaction} did not give any additional improvements to the unshifted interaction picture, following the same rationale as the previously proposed symmetry shift. For a more detailed discussion, see Ref.~\citenum{loaiza_lcu}. 

\section{Conclusions}
We have presented the \gls{BLISS} methodology for more efficient encodings of the moleculular electronic structure Hamiltonians. The key idea of \gls{BLISS} is the modification of $\hat H$ that reduces its spectral range and 1-norm LCU cost without changing its action in the Fock subspace of a particular molecular symmetry. By substituting $\hat H$ with a modified operator $\hat H_T$ in the energy estimation algorithm lowers the overall cost without affecting results. The target number of electrons, which is used when constructing $\hat H_T$, is specified when an initial wavefunction is considered in the energy estimation algorithm. \blue{Alternatively, even if the initial wavefunction is not an eigenstate of the electron number operator, the ground state energy can still be recovered as long as the overlap with the ground state is larger than the overlap with any other eigenstate.} The \gls{BLISS} methodology yields a shifted Hamiltonian $\hat H_T$ of the same operator form as the original two-electron Hamiltonian $\hat H$ [Eq.\eqref{eq:mol_ham}], the only difference is in modified electron-integral coefficients. As such, this methodology can be used as a preprocessing step for any electronic structure Hamiltonian, and can be thought as obtaining a ``new molecule'' that is isospectral to $\hat H$ in the subspace with a target number of electrons. \blue{The $2.2\%$ average decrease in the required number of unitaries for the LCU decompositions after the symmetry shift means that the only factor modifying the cost of the phase estimation algorithm will be the change in 1-norm of the shifted Hamiltonian.}

Application of the \gls{BLISS} reduced 1-norms of LCU encodings by a factor $\sim 2$ and spectral \blue{ranges} of electronic Hamiltonians by a factor of $\sim 1.7$. This practically halves the cost of energy estimation routines on a quantum computer.

\section*{Acknowledgements}
I.L. and A.F.I. would like to thank Marcel Nooijen, Luis A. Martinez Martinez, and Guoming Wang for stimulating discussions. I.L. and A.F.I. gratefully appreciate financial support from the Mitacs Elevate Postdoctoral Fellowship and Zapata Computing Inc. 

\appendix
\section{Appendix A: LCU decompositions} \label{app:lcu}
Here we review the different \gls{LCU} decompositions that were used in this work.

\subsubsection{Qubit-based approaches}
The simplest \gls{LCU} decomposition is obtained when $\hat H$ is mapped into a qubit operator by means of a fermion-to-qubit mapping, e.g. Jordan-Wigner \cite{jordan_wigner} or Bravyi-Kitaev \cite{bk1,bk2,bk3}. This yields the expression
\begin{equation} \label{eq:H_q}
    \hat H = \sum_q d_q \hat P_q,
\end{equation}
where $d_q$ are constants and $\hat P_q$ are Pauli-product operators consisting of products of Pauli matrices acting on different qubits. Given how $\hat P_q$'s are already unitary, this expression is already an \gls{LCU} of the Hamiltonian with the 1-norm
\begin{equation}
    \lambda^{(P)} = \sum_q |d_q|.
\end{equation}
We note that by expressing the Hamiltonian using Majorana operators, this 1-norm can also be obtained as a function of the fermionic tensors \cite{majorana_l1,loaiza_lcu}:
\begin{align} \label{eq:l1_majorana}
    \lambda^{(P)} &= \sum_{ij} |h_{ij} + 2\sum_k g_{ijkk}| + \sum_{i>k,j>l} |g_{ijkl} - g_{ilkj}| \nonumber \\
    &\ + \frac{1}{2}\sum_{ijkl} |g_{ijkl}|.
\end{align}
When starting from the qubit Hamiltonian [Eq.\eqref{eq:H_q}], there are two additional optimizations that can be done for lowering the encoding cost. The orbital optimization technique \cite{majorana_l1} applies an orbital rotation
\begin{equation} \label{eq:orb_rot}
    \hat U(\vec\theta) \equiv \prod_{i>j} e^{\theta_{ij} (\hat F^i_j - \hat F_i^j)},
\end{equation}
finding the optimal parameters $\vec \theta$ such that the 1-norm of $\hat U(\vec\theta) \hat H \hat U^\dagger(\vec\theta)$ is minimized. In addition, we can also make use of the anticommuting grouping technique \cite{anticommuting,loaiza_lcu}. The key idea of this approach is that a normalized linear combination of mutually anticommuting Pauli products yields a unitary operator. Noting that two arbitrary Pauli products $\hat P_i$ and $\hat P_j$ always either commute ($[\hat P_i,\hat P_j] = 0$) or anticommute ($\{\hat P_i,\hat P_j\} = 0$), mutually anticommuting Pauli products are grouped into sets with corresponding indices $G_n$. We thus obtain the \gls{LCU}
\begin{equation}
    \hat H = \sum_n a_n \hat A_n,
\end{equation}
where $\hat A_n \equiv \frac{1}{a_n}\sum_{q\in G_n} d_q \hat P_q$ are unitary operators and $a_n \equiv \sqrt{\sum_{q\in G_n} |d_q|^2 }$. It can be easily shown \cite{loaiza_lcu} that the resulting 1-norm of this decomposition,
\begin{equation}
    \lambda^{(AC)} = \sum_n |a_n|
\end{equation}
is always smaller than $\lambda^{(P)}$ if there is a nontrivial grouping. 

\subsubsection{Fermionic-based approaches}
We now give an overview of fermionic-based approaches. We start by giving the \gls{CSA} form of the Hamiltonian \cite{CSA, parrish_csa, csa_constrained}:
\begin{align}
    \hat H_1 &\equiv \sum_{ij} h_{ij} \hat F^i_j \\
    &= \hat U_1 \left(\sum_i \lambda_{ii}^{(1)} \sum_\sigma\hat n_{i\sigma} \right) \hat U_1^\dagger,
\end{align}
and
\begin{align}
    \hat H_2 &\equiv \sum_{ijkl} g_{ijkl} \hat F^i_j \hat F^k_l \\
    &= \sum_{m\geq 2}\hat U_m \left(\sum_{ij} \lambda_{ij}^{(m)} \sum_{\sigma\tau}\hat n_{i\sigma} \hat n_{j\tau} \right) \hat U_m^\dagger,
\end{align}
where we have defined the one(two)-electron Hamiltonians $\hat H_{1(2)}$ such that $\hat H = \hat H_1 + \hat H_2$, $\{\sigma,\tau\}$ correspond to $z$-spin projections $\{\alpha,\beta\}$, $\hat n_{i\sigma} \equiv \hat a^\dagger_{i\sigma}\hat a_{i\sigma}$ is the number operator on orbital $i$ with spin $\sigma$, and $\hat U_m$'s are orbital rotations as seen in Eq.\eqref{eq:orb_rot}.  By noting that the number operators $\hat n_{i\sigma}$ can be mapped into reflections, and thus unitaries, by the transformation $\hat n_{i\sigma} \rightarrow 2\hat n_{i\sigma} - \hat 1$. Defining the reflections $\hat r_{i\sigma} \equiv 2\hat n_{i\sigma} - \hat 1$, the \gls{CSA} form of the Hamiltonian can be turned into an \gls{LCU} as
\begin{equation}
    \hat H_2^{\textrm{(LCU)}} = \sum_m \hat U_m \left( \sum_{ij} \frac{\lambda_{ij}^{(m)}}{4} \sum_{\sigma\tau} \hat r_{i\sigma} \hat r_{j\tau} \right) \hat U_m^\dagger.
\end{equation}
An adjustment of $\hat H_1$ must be done to account for the one-electron terms coming from the $\hat n_{i\sigma}\rightarrow\hat r_{i\sigma}$ mapping, which can be shown to yield \cite{THC}
\begin{equation}
    \hat H_1' \equiv \hat H_1 + 2 \sum_{ijk} g_{ijkk} \hat F^i_j.
\end{equation}
This one-electron operator can then be diagonalized as $\hat H_1' = \hat U_1' \left( \sum_i \mu_i \sum_\sigma \hat n_{i\sigma} \right) \hat U_1'^\dagger$, obtaining the \gls{LCU}
\begin{equation}
    \hat H_1^{(\textrm{LCU})} = \hat U_1' \left( \sum_i \frac{\mu_i}{2} \sum_\sigma \hat r_{i\sigma} \right) \hat U_1'^\dagger.
\end{equation}
The resulting 1-norm of this decomposition then corresponds to
\begin{align}
    \lambda^{(\textrm{CSA})} &= \lambda_1 + \lambda_2^{(\textrm{CSA})} \nonumber \\
    &= \sum_i |\mu_i| + \sum_{m\geq 2} \sum_{ij} |\lambda^{(m)}_{ij}| - \frac{1}{2}\sum_{m\geq 2} \sum_i |\lambda^{(m)}_{ii}|,
\end{align}
where we have removed some operators associated with the $\lambda_{ii}^{(m)}$ coefficients due to the fact that $\hat r_{i\sigma}^2 = \hat 1$. Note that the modified one-electron fragment is independent of the fermionic \gls{LCU} decomposition method. 

An alternative \gls{LCU} decomposition can be made through the \gls{DF} decomposition \cite{femoco_df,qubitized_pe, df_1, df_2, df_3, df_4, df_5}, which can be considered as a \gls{CSA} decomposition with a rank-deficient tensor $\lambda_{ij}^{(m)} = \epsilon^{(m)}_i \epsilon^{(m)}_j$. This allows for the Hamiltonian to be written as
\begin{equation}
    \hat H_2 = \sum_m \hat U_m \left( \sum_i \epsilon_i^{(m)} \sum_\sigma \hat n_{i\sigma} \right)^2 \hat U_m^\dagger.
\end{equation}
The complete-square structure of each fragment then allows for the implementation of each $m$-th fragment as a single unitary through the use of a qubitization procedure \cite{qubitized_pe,THC}, yielding a correspoding 1-norm of
\begin{align}
    \lambda^{(\textrm{DF})} &= \lambda_1 + \lambda_2^{(\textrm{DF})} \\
    &= \sum_i |\mu_i| + \frac{1}{2} \sum_{m\geq 2}\left( \sum_i \epsilon_i^{(m)}\right)^2.
\end{align}
For a more detailed discussion of how the \gls{DF} \gls{LCU} is implemented, we refer to Refs.~\citenum{femoco_df,loaiza_lcu,THC}. 

Finally, we note that we have skipped the Tensor Hypercontraction method \cite{THC,bloch_thc} for \gls{LCU} decompositions. This is due to the current optimization procedures for this decomposition being unstable and having a poor convergence to $\hat H$: the non-linear nature of this \textit{ansatz} makes the optimization extremely sensitive to initial conditions and to the chosen optimization method. Work for obtaining this decomposition in a robust and stable way is under progress. 

\section{Appendix B: Molecular geometries and computational details} \label{app:comp}
In this section we write down all details necessary for numerical reproducibility. Our code is available at \href{https://github.com/iloaiza/QuantumMAMBO.jl}{https://github.com/iloaiza/QuantumMAMBO.jl}. 

For the CSA decompositions of the two-electron tensor, the cost function was chosen as the 2-norm of $\Delta \hat H \equiv \sum_{ijkl} b_{ijkl}\hat F^{i}_{j}\hat F^{k}_{l}$:
\begin{equation}
    ||\Delta\hat H||_2 \equiv \sum_{ijkl} |b_{ijkl}|^2,
\end{equation}
considering the decomposition finished when this norm is below a tolerance of $1\times 10^{-6}$. All non-linear optimizations where done using the Julia Optim.jl package \cite{optim}, using the BFGS algorithm \cite{bfgs} with the default tolerance. Linear programming routines were done using the Julia JuMP package \cite{jump} with the HiGHS optimizer \cite{highs}.

All molecular Hamiltonians were generated using the PySCF package \cite{pyscf1,pyscf2,pyscf3} and the Openfermion library \cite{openfermion}, using a minimal STO-3G basis \cite{sto3g,szabo} and the Jordan-Wigner transformation \cite{jordan_wigner}. The nuclear geometries for the Hamiltonians are:
\begin{itemize}
    \item R(H -- H) = $1\textrm{\AA}$ for H$_2$
    \item R(Li -- H) = $1\textrm{\AA}$ for LiH
    \item R(Be -- H) = $1\textrm{\AA}$ with a collinear atomic arrangement for BeH$_2$
    \item R(O -- H) = $1\textrm{\AA}$ with angle $\angle$HOH = $107.6^\circ$ for H$_2$O
    \item R(N -- H) = $1\textrm{\AA}$ with $\angle$HNH = $107^\circ$ for NH$_3$
\end{itemize}

\bibliography{biblio}
\end{document}

%% file: tables.tex
\begin{sidewaystable*}[]
\centering
\vspace{7cm}
\begin{tabular}{|c|c|c|c|c|c|c|c|@{\hskip 0.1cm}|@{\hskip 0.1cm}|c|c|}
\hline
System                               & Hamiltonian  & Pauli        & OO-Pauli      & AC           & OO-AC        & DF         & GCSA &  $\Delta E/2$  & \blue{$\Delta E_{N_e}/2$} \\ \hline
\multirow{3}{*}{H$_2$}               & $\hat H$              & 1.58(22) & 1.58(22) & 1.49(18) & 1.49(18) & 1.37(7) & 1.77(20) & 0.815  & \multirow{3}{*}{\blue{0.57}} \\
                                     & $\hat H_S$             & 0.842(18) & 0.842(18) & 0.795(16) & 0.795(16) & 0.741(5) & 0.842(16) & 0.656  &  \\
                                     & $\hat H_T$             & 0.839(18) & 0.839(18) & 0.75(14) & 0.75(14) & 0.741(7) & 0.839(16) & 0.57   &     \\ \hline
\multirow{3}{*}{LiH}                 & $\hat H$              & 13.0(1086) & 12.4(1058) & 10.2(168) & 10.2(164) & 9.34(33) & 11.0(1434) & 4.93  & \multirow{3}{*}{\blue{3.52}}      \\
                                     & $\hat H_S$             & 7.62(1082) & 7.02(1072) & 5.13(160) & 5.03(156) & 4.76(31) & 5.45(1212)  & 3.57   &     \\
                                     & $\hat H_T$             & 6.98(1050) & 6.3(1070) & 4.86(176) & 4.67(168) & 4.64(33) & 5.21(1374)  & 3.53   &  \\ \hline
\multirow{3}{*}{BeH$_2$}             & $\hat H$           & 22.8(1142) & 21.9(1122) & 18.0(198) & 17.9(192) & 16.4(42) & 20.6(2420) & 9.99  & \multirow{3}{*}{\blue{7.29}}  \\
                                     & $\hat H_S$             & 14.2(1134) & 13.0(1128) & 10.2(194) & 9.86(188) & 9.77(38) & 11.7(2230)    & 7.31   &  \\
                                     & $\hat H_T$             & 13.2(1126) & 12.0(1138) & 9.6(198) & 9.18(196) & 9.55(42) & 10.8(2532)   & 7.35  &    \\ \hline
\multirow{3}{*}{H$_2$O}              & $\hat H$            & 71.9(1862) & 61.0(1838) & 57.2(240) & 55.7(244) & 53.7(42) & 58.9(3042)  & 41.9   & \multirow{3}{*}{\blue{23.7}}   \\
                                     & $\hat H_S$             & 46.0(1858) & 37.7(1846) & 34.4(232) & 32.9(246) & 32.7(40) & 36.1(2934)  & 28.9 &  \\
                                     & $\hat H_T$             & 35.5(1862) & 31.3(1806) & 27.9(230) & 27.0(240) & 27.6(42) & 29.9(3034) & 23.8 &  \\ \hline
\multirow{3}{*}{NH$_3$}              &$\hat H$              & 70.6(6280) & 54.5(3880) & 49.1(528) & 46.8(464) & 44.7(52) & 50.6(5252)  & 33.8  & \multirow{3}{*}{\blue{19.5}} \\
                                     & $\hat H_S$             & 48.2(6272) & 34.6(3864) & 30.1(520) & 27.8(436) & 28.1(50) & 43.3(7298) & 23.1 &  \\ 
                                     & $\hat H_T$             & 38.7(6268) & 30.8(4036) & 25.3(602) & 24.1(534) & 24.9(52) & 27.1(5364)   &19.8 &  \\ \hline \hline
\multirow{3}{*}{Linear fit slope} & $\hat H$           & 1      & 0.81$\pm$0.02 & 0.75$\pm$0.03 & 0.72$\pm$0.03 & \textbf{0.69$\pm$0.03} & 0.78$\pm$0.03 & 1 & \multirow{3}{*}{\blue{0.58 $\pm$ 0.016}} \\
                                              & $\hat H_S$   & 0.66$\pm$0.01 & 0.51$\pm$0.01 & 0.45$\pm$0.01 & \textbf{0.43$\pm$0.02} & \textbf{0.43$\pm$0.01} & 0.55$\pm$0.03 & 0.69$\pm$0.005 & \\
                                              & $\hat H_T$    & 0.52$\pm$0.02 & 0.44$\pm$0.01  & 0.38$\pm$0.01 & \textbf{0.36$\pm$0.01} & 0.37$\pm$0.01 & 0.40$\pm$0.01 & 0.58$\pm$0.016 & \\ \hline

\end{tabular}
\caption{1-norms for molecular Hamiltonian using different LCU decompositions: 
$\Delta E/2 = (E_{\rm max}-E_{\rm min})/2$ is the spectral range and a lower bound to the 1-norm; $\Delta E_{N_e}/2$ is the spectral range in the subspace with $N_e$ electrons and is invariant to symmetry shifts; Pauli, Pauli products; AC, anticommuting Pauli product grouping; OO-Pauli, Pauli products with orbital optimization for 1-norm \cite{majorana_l1}; OO-AC, orbital optimization scheme with subsequent anticommuting Pauli product grouping \cite{majorana_l1,loaiza_lcu}; DF, double factorization; GCSA, Greedy CSA decomposition. 
Values in parenthesis represent the total number of unitaries in the \gls{LCU}, which is associated with the necessary number of ancilla qubits and implementation cost (see Refs.\citenum{THC,femoco_df} for a more detailed discussion). Hamiltonians correspond to Eqs.(\ref{eq:HS},\ref{eq:bliss}). A cut-off threshold of $10^{-6}$ was used for counting unitaries. The linear fit was obtained by associating to each molecule two coordinates: $x$ is the 1-norm of the Pauli product LCU, and $y$ is the 1-norm of the considered LCU method.  The slope of the linear regression is given with the associated standard error ($\pm$). Highlights show methods with best scaling. For $\Delta E/2$ and $\Delta E_{N_e}/2$, the linear fit column corresponds to the average improvement with respect to the unshifted $\Delta E/2$. Note that each of the DF unitaries requires an implementation via qubitization, as such it is difficult to compare the full cost of implementing the resulting Hamiltonian oracle with that of other methods without explicit construction of the oracle circuits.}
\label{tab:results}
\end{sidewaystable*}

\begin{table}
    \centering
    \begin{tabular}{|c|c|c|c||c|c|c|} \hline
      & $\alpha[\hat H]$    & $\alpha[\hat H_S]$ & $\alpha[\hat H_T]$ & $\beta[\hat H]$    & $\beta[\hat H_S]$ & $\beta[\hat H_T]$  \\ \hline
      Pauli (CMOs) & 2.25 & 2.34 & 2.35 & -0.57 & -0.73 & -0.75 \\ \hline
      Pauli (FB) & 1.25 & 1.37 & 1.44 & -0.12 & -0.29 & -0.34 \\ \hline
      AC (CMOs) & 1.68 & 1.81 & 1.81 & -0.43 & -0.67 & -0.68 \\ \hline
      AC (FB) & 1.30 & 1.43 & 1.47 & -0.28 & -0.49 & -0.54 \\ \hline
      DF & 1.91 & 2.07 & 2.05 & -0.61 & -0.89 & -0.88 \\ \hline
    \end{tabular}
    \caption{\blue{Fitting coefficients for Fig.~\ref{fig:scaling}.  $R^2$ regression coefficient for all curves is between $0.995$ and $1$ and is not reported.}} 
    \label{tab:scaling}
\end{table}

%% file: main.bbl
\begin{thebibliography}{62}%
\makeatletter
\providecommand \@ifxundefined [1]{%
 \@ifx{#1\undefined}
}%
\providecommand \@ifnum [1]{%
 \ifnum #1\expandafter \@firstoftwo
 \else \expandafter \@secondoftwo
 \fi
}%
\providecommand \@ifx [1]{%
 \ifx #1\expandafter \@firstoftwo
 \else \expandafter \@secondoftwo
 \fi
}%
\providecommand \natexlab [1]{#1}%
\providecommand \enquote  [1]{``#1''}%
\providecommand \bibnamefont  [1]{#1}%
\providecommand \bibfnamefont [1]{#1}%
\providecommand \citenamefont [1]{#1}%
\providecommand \href@noop [0]{\@secondoftwo}%
\providecommand \href [0]{\begingroup \@sanitize@url \@href}%
\providecommand \@href[1]{\@@startlink{#1}\@@href}%
\providecommand \@@href[1]{\endgroup#1\@@endlink}%
\providecommand \@sanitize@url [0]{\catcode `\\12\catcode `\$12\catcode
  `\&12\catcode `\#12\catcode `\^12\catcode `\_12\catcode `\%12\relax}%
\providecommand \@@startlink[1]{}%
\providecommand \@@endlink[0]{}%
\providecommand \url  [0]{\begingroup\@sanitize@url \@url }%
\providecommand \@url [1]{\endgroup\@href {#1}{\urlprefix }}%
\providecommand \urlprefix  [0]{URL }%
\providecommand \Eprint [0]{\href }%
\providecommand \doibase [0]{http://dx.doi.org/}%
\providecommand \selectlanguage [0]{\@gobble}%
\providecommand \bibinfo  [0]{\@secondoftwo}%
\providecommand \bibfield  [0]{\@secondoftwo}%
\providecommand \translation [1]{[#1]}%
\providecommand \BibitemOpen [0]{}%
\providecommand \bibitemStop [0]{}%
\providecommand \bibitemNoStop [0]{.\EOS\space}%
\providecommand \EOS [0]{\spacefactor3000\relax}%
\providecommand \BibitemShut  [1]{\csname bibitem#1\endcsname}%
\let\auto@bib@innerbib\@empty
\bibitem [{\citenamefont {Cao}\ \emph {et~al.}(2019)\citenamefont {Cao},
  \citenamefont {Romero}, \citenamefont {Olson}, \citenamefont {Degroote},
  \citenamefont {Johnson}, \citenamefont {Kieferová}, \citenamefont
  {Kivlichan}, \citenamefont {Menke}, \citenamefont {Peropadre}, \citenamefont
  {Sawaya}, \citenamefont {Sim}, \citenamefont {Veis},\ and\ \citenamefont
  {Aspuru-Guzik}}]{QCinQC}%
  \BibitemOpen
  \bibfield  {author} {\bibinfo {author} {\bibfnamefont {Y.}~\bibnamefont
  {Cao}}, \bibinfo {author} {\bibfnamefont {J.}~\bibnamefont {Romero}},
  \bibinfo {author} {\bibfnamefont {J.~P.}\ \bibnamefont {Olson}}, \bibinfo
  {author} {\bibfnamefont {M.}~\bibnamefont {Degroote}}, \bibinfo {author}
  {\bibfnamefont {P.~D.}\ \bibnamefont {Johnson}}, \bibinfo {author}
  {\bibfnamefont {M.}~\bibnamefont {Kieferová}}, \bibinfo {author}
  {\bibfnamefont {I.~D.}\ \bibnamefont {Kivlichan}}, \bibinfo {author}
  {\bibfnamefont {T.}~\bibnamefont {Menke}}, \bibinfo {author} {\bibfnamefont
  {B.}~\bibnamefont {Peropadre}}, \bibinfo {author} {\bibfnamefont {N.~P.~D.}\
  \bibnamefont {Sawaya}}, \bibinfo {author} {\bibfnamefont {S.}~\bibnamefont
  {Sim}}, \bibinfo {author} {\bibfnamefont {L.}~\bibnamefont {Veis}}, \ and\
  \bibinfo {author} {\bibfnamefont {A.}~\bibnamefont {Aspuru-Guzik}},\ }\href
  {https://doi.org/10.1021/acs.chemrev.8b00803} {\bibfield  {journal} {\bibinfo
   {journal} {Chem. Rev.}\ }\textbf {\bibinfo {volume} {119}},\ \bibinfo
  {pages} {10856} (\bibinfo {year} {2019})}\BibitemShut {NoStop}%
\bibitem [{\citenamefont {Kassal}\ \emph {et~al.}(2008)\citenamefont {Kassal},
  \citenamefont {Jordan}, \citenamefont {Love}, \citenamefont {Mohseni},\ and\
  \citenamefont {Aspuru-Guzik}}]{polynomial_dynamics}%
  \BibitemOpen
  \bibfield  {author} {\bibinfo {author} {\bibfnamefont {I.}~\bibnamefont
  {Kassal}}, \bibinfo {author} {\bibfnamefont {S.~P.}\ \bibnamefont {Jordan}},
  \bibinfo {author} {\bibfnamefont {P.~J.}\ \bibnamefont {Love}}, \bibinfo
  {author} {\bibfnamefont {M.}~\bibnamefont {Mohseni}}, \ and\ \bibinfo
  {author} {\bibfnamefont {A.}~\bibnamefont {Aspuru-Guzik}},\ }\href
  {https://www.pnas.org/doi/abs/10.1073/pnas.0808245105} {\bibfield  {journal}
  {\bibinfo  {journal} {Proc. Natl. Acad. Sci.}\ }\textbf {\bibinfo {volume}
  {105}},\ \bibinfo {pages} {18681} (\bibinfo {year} {2008})}\BibitemShut
  {NoStop}%
\bibitem [{\citenamefont {Abrams}\ and\ \citenamefont {Lloyd}(1999)}]{QPE}%
  \BibitemOpen
  \bibfield  {author} {\bibinfo {author} {\bibfnamefont {D.~S.}\ \bibnamefont
  {Abrams}}\ and\ \bibinfo {author} {\bibfnamefont {S.}~\bibnamefont {Lloyd}},\
  }\href {\doibase 10.1103/PhysRevLett.83.5162} {\bibfield  {journal} {\bibinfo
   {journal} {Phys. Rev. Lett.}\ }\textbf {\bibinfo {volume} {83}},\ \bibinfo
  {pages} {5162} (\bibinfo {year} {1999})}\BibitemShut {NoStop}%
\bibitem [{\citenamefont {Kitaev}(1995)}]{kitaev_QPE}%
  \BibitemOpen
  \bibfield  {author} {\bibinfo {author} {\bibfnamefont {A.~Y.}\ \bibnamefont
  {Kitaev}},\ }\href {https://arxiv.org/abs/quant-ph/9511026} {\enquote
  {\bibinfo {title} {Quantum measurements and the {Abelian} stabilizer
  problem},}\ } (\bibinfo {year} {1995}),\ \Eprint
  {http://arxiv.org/abs/9511026} {arXiv:9511026 [quant-ph]} \BibitemShut
  {NoStop}%
\bibitem [{\citenamefont {Aspuru-Guzik}\ \emph {et~al.}(2005)\citenamefont
  {Aspuru-Guzik}, \citenamefont {Dutoi}, \citenamefont {Love},\ and\
  \citenamefont {Head-Gordon}}]{aspuru_qpe}%
  \BibitemOpen
  \bibfield  {author} {\bibinfo {author} {\bibfnamefont {A.}~\bibnamefont
  {Aspuru-Guzik}}, \bibinfo {author} {\bibfnamefont {A.~D.}\ \bibnamefont
  {Dutoi}}, \bibinfo {author} {\bibfnamefont {P.~J.}\ \bibnamefont {Love}}, \
  and\ \bibinfo {author} {\bibfnamefont {M.}~\bibnamefont {Head-Gordon}},\
  }\href {https://doi.org/10.1126%2Fscience.1113479} {\bibfield  {journal}
  {\bibinfo  {journal} {Science}\ }\textbf {\bibinfo {volume} {309}},\ \bibinfo
  {pages} {1704} (\bibinfo {year} {2005})}\BibitemShut {NoStop}%
\bibitem [{\citenamefont {Lin}\ and\ \citenamefont
  {Tong}(2022)}]{heisenberg_qpe}%
  \BibitemOpen
  \bibfield  {author} {\bibinfo {author} {\bibfnamefont {L.}~\bibnamefont
  {Lin}}\ and\ \bibinfo {author} {\bibfnamefont {Y.}~\bibnamefont {Tong}},\
  }\href {https://doi.org/10.1103%2Fprxquantum.3.010318} {\bibfield  {journal}
  {\bibinfo  {journal} {{PRX} Quantum}\ }\textbf {\bibinfo {volume} {3}},\
  \bibinfo {pages} {010318} (\bibinfo {year} {2022})}\BibitemShut {NoStop}%
\bibitem [{\citenamefont {Moore}\ \emph {et~al.}(2021)\citenamefont {Moore},
  \citenamefont {Wang}, \citenamefont {Hu}, \citenamefont {Kais},\ and\
  \citenamefont {Weiner}}]{statistical_qpe}%
  \BibitemOpen
  \bibfield  {author} {\bibinfo {author} {\bibfnamefont {A.~J.}\ \bibnamefont
  {Moore}}, \bibinfo {author} {\bibfnamefont {Y.}~\bibnamefont {Wang}},
  \bibinfo {author} {\bibfnamefont {Z.}~\bibnamefont {Hu}}, \bibinfo {author}
  {\bibfnamefont {S.}~\bibnamefont {Kais}}, \ and\ \bibinfo {author}
  {\bibfnamefont {A.~M.}\ \bibnamefont {Weiner}},\ }\href
  {https://doi.org/10.1088%2F1367-2630%2Fac320d} {\bibfield  {journal}
  {\bibinfo  {journal} {New J. Phys.}\ }\textbf {\bibinfo {volume} {23}},\
  \bibinfo {pages} {113027} (\bibinfo {year} {2021})}\BibitemShut {NoStop}%
\bibitem [{\citenamefont {Somma}(2019)}]{ts_qpe}%
  \BibitemOpen
  \bibfield  {author} {\bibinfo {author} {\bibfnamefont {R.~D.}\ \bibnamefont
  {Somma}},\ }\href {https://dx.doi.org/10.1088/1367-2630/ab5c60} {\bibfield
  {journal} {\bibinfo  {journal} {New J. Phys.}\ }\textbf {\bibinfo {volume}
  {21}},\ \bibinfo {pages} {123025} (\bibinfo {year} {2019})}\BibitemShut
  {NoStop}%
\bibitem [{\citenamefont {Ge}\ \emph {et~al.}(2019)\citenamefont {Ge},
  \citenamefont {Tura},\ and\ \citenamefont {Cirac}}]{cirac_qpe}%
  \BibitemOpen
  \bibfield  {author} {\bibinfo {author} {\bibfnamefont {Y.}~\bibnamefont
  {Ge}}, \bibinfo {author} {\bibfnamefont {J.}~\bibnamefont {Tura}}, \ and\
  \bibinfo {author} {\bibfnamefont {J.~I.}\ \bibnamefont {Cirac}},\ }\href
  {https://doi.org/10.1063/1.5027484} {\bibfield  {journal} {\bibinfo
  {journal} {J. Math. Phys.}\ }\textbf {\bibinfo {volume} {60}},\ \bibinfo
  {pages} {022202} (\bibinfo {year} {2019})}\BibitemShut {NoStop}%
\bibitem [{\citenamefont {Wang}\ \emph {et~al.}(2022)\citenamefont {Wang},
  \citenamefont {Stilck-França}, \citenamefont {Zhang}, \citenamefont {Zhu},\
  and\ \citenamefont {Johnson}}]{GSEE}%
  \BibitemOpen
  \bibfield  {author} {\bibinfo {author} {\bibfnamefont {G.}~\bibnamefont
  {Wang}}, \bibinfo {author} {\bibfnamefont {D.}~\bibnamefont
  {Stilck-França}}, \bibinfo {author} {\bibfnamefont {R.}~\bibnamefont
  {Zhang}}, \bibinfo {author} {\bibfnamefont {S.}~\bibnamefont {Zhu}}, \ and\
  \bibinfo {author} {\bibfnamefont {P.~D.}\ \bibnamefont {Johnson}},\ }\href
  {https://arxiv.org/abs/2209.06811} {\enquote {\bibinfo {title} {Quantum
  algorithm for ground state energy estimation using circuit depth with
  exponentially improved dependence on precision},}\ } (\bibinfo {year}
  {2022})\BibitemShut {NoStop}%
\bibitem [{\citenamefont {Childs}\ and\ \citenamefont {Wiebe}(2012)}]{LCU}%
  \BibitemOpen
  \bibfield  {author} {\bibinfo {author} {\bibfnamefont {A.~M.}\ \bibnamefont
  {Childs}}\ and\ \bibinfo {author} {\bibfnamefont {N.}~\bibnamefont {Wiebe}},\
  }\href {https://dl.acm.org/doi/10.5555/2481569.2481570} {\bibfield  {journal}
  {\bibinfo  {journal} {Quantum Info. Comput.}\ }\textbf {\bibinfo {volume}
  {12}},\ \bibinfo {pages} {901–924} (\bibinfo {year} {2012})}\BibitemShut
  {NoStop}%
\bibitem [{\citenamefont {Low}\ and\ \citenamefont
  {Wiebe}(2019)}]{interaction}%
  \BibitemOpen
  \bibfield  {author} {\bibinfo {author} {\bibfnamefont {G.~H.}\ \bibnamefont
  {Low}}\ and\ \bibinfo {author} {\bibfnamefont {N.}~\bibnamefont {Wiebe}},\
  }\href {https:/arXiv.org/abs/1805.00675} {\enquote {\bibinfo {title}
  {{Hamiltonian} simulation in the interaction picture},}\ } (\bibinfo {year}
  {2019}),\ \Eprint {http://arxiv.org/abs/1805.00675} {arXiv:1805.00675
  [quant-ph]} \BibitemShut {NoStop}%
\bibitem [{\citenamefont {Berry}\ \emph {et~al.}(2019)\citenamefont {Berry},
  \citenamefont {Gidney}, \citenamefont {Motta}, \citenamefont {McClean},\ and\
  \citenamefont {Babbush}}]{qubitized_pe}%
  \BibitemOpen
  \bibfield  {author} {\bibinfo {author} {\bibfnamefont {D.~W.}\ \bibnamefont
  {Berry}}, \bibinfo {author} {\bibfnamefont {C.}~\bibnamefont {Gidney}},
  \bibinfo {author} {\bibfnamefont {M.}~\bibnamefont {Motta}}, \bibinfo
  {author} {\bibfnamefont {J.~R.}\ \bibnamefont {McClean}}, \ and\ \bibinfo
  {author} {\bibfnamefont {R.}~\bibnamefont {Babbush}},\ }\href
  {https://doi.org/10.22331%2Fq-2019-12-02-208} {\bibfield  {journal} {\bibinfo
   {journal} {Quantum}\ }\textbf {\bibinfo {volume} {3}},\ \bibinfo {pages}
  {208} (\bibinfo {year} {2019})}\BibitemShut {NoStop}%
\bibitem [{\citenamefont {Lee}\ \emph {et~al.}(2021)\citenamefont {Lee},
  \citenamefont {Berry}, \citenamefont {Gidney}, \citenamefont {Huggins},
  \citenamefont {McClean}, \citenamefont {Wiebe},\ and\ \citenamefont
  {Babbush}}]{THC}%
  \BibitemOpen
  \bibfield  {author} {\bibinfo {author} {\bibfnamefont {J.}~\bibnamefont
  {Lee}}, \bibinfo {author} {\bibfnamefont {D.~W.}\ \bibnamefont {Berry}},
  \bibinfo {author} {\bibfnamefont {C.}~\bibnamefont {Gidney}}, \bibinfo
  {author} {\bibfnamefont {W.~J.}\ \bibnamefont {Huggins}}, \bibinfo {author}
  {\bibfnamefont {J.~R.}\ \bibnamefont {McClean}}, \bibinfo {author}
  {\bibfnamefont {N.}~\bibnamefont {Wiebe}}, \ and\ \bibinfo {author}
  {\bibfnamefont {R.}~\bibnamefont {Babbush}},\ }\href
  {https://doi.org/10.1103%2Fprxquantum.2.030305} {\bibfield  {journal}
  {\bibinfo  {journal} {{PRX} Quantum}\ }\textbf {\bibinfo {volume} {2}},\
  \bibinfo {pages} {030305} (\bibinfo {year} {2021})}\BibitemShut {NoStop}%
\bibitem [{\citenamefont {Suzuki}(1991)}]{trotter}%
  \BibitemOpen
  \bibfield  {author} {\bibinfo {author} {\bibfnamefont {M.}~\bibnamefont
  {Suzuki}},\ }\href {https://doi.org/10.1063/1.529425} {\bibfield  {journal}
  {\bibinfo  {journal} {J. Math. Phys.}\ }\textbf {\bibinfo {volume} {32}},\
  \bibinfo {pages} {400} (\bibinfo {year} {1991})}\BibitemShut {NoStop}%
\bibitem [{\citenamefont {Low}\ and\ \citenamefont
  {Chuang}(2019)}]{low2019hamiltonian}%
  \BibitemOpen
  \bibfield  {author} {\bibinfo {author} {\bibfnamefont {G.~H.}\ \bibnamefont
  {Low}}\ and\ \bibinfo {author} {\bibfnamefont {I.~L.}\ \bibnamefont
  {Chuang}},\ }\href {https://doi.org/10.22331/q-2019-07-12-163} {\bibfield
  {journal} {\bibinfo  {journal} {Quantum}\ }\textbf {\bibinfo {volume} {3}},\
  \bibinfo {pages} {163} (\bibinfo {year} {2019})}\BibitemShut {NoStop}%
\bibitem [{\citenamefont {Babbush}\ \emph {et~al.}(2018)\citenamefont
  {Babbush}, \citenamefont {Gidney}, \citenamefont {Berry}, \citenamefont
  {Wiebe}, \citenamefont {McClean}, \citenamefont {Paler}, \citenamefont
  {Fowler},\ and\ \citenamefont {Neven}}]{qubitization_wiebe}%
  \BibitemOpen
  \bibfield  {author} {\bibinfo {author} {\bibfnamefont {R.}~\bibnamefont
  {Babbush}}, \bibinfo {author} {\bibfnamefont {C.}~\bibnamefont {Gidney}},
  \bibinfo {author} {\bibfnamefont {D.~W.}\ \bibnamefont {Berry}}, \bibinfo
  {author} {\bibfnamefont {N.}~\bibnamefont {Wiebe}}, \bibinfo {author}
  {\bibfnamefont {J.}~\bibnamefont {McClean}}, \bibinfo {author} {\bibfnamefont
  {A.}~\bibnamefont {Paler}}, \bibinfo {author} {\bibfnamefont
  {A.}~\bibnamefont {Fowler}}, \ and\ \bibinfo {author} {\bibfnamefont
  {H.}~\bibnamefont {Neven}},\ }\href
  {https://link.aps.org/doi/10.1103/PhysRevX.8.041015} {\bibfield  {journal}
  {\bibinfo  {journal} {Phys. Rev. X}\ }\textbf {\bibinfo {volume} {8}},\
  \bibinfo {pages} {041015} (\bibinfo {year} {2018})}\BibitemShut {NoStop}%
\bibitem [{\citenamefont {Delgado}\ \emph {et~al.}(2022)\citenamefont
  {Delgado}, \citenamefont {Casares}, \citenamefont {dos Reis}, \citenamefont
  {Zini}, \citenamefont {Campos}, \citenamefont {Cruz-Hern\'andez},
  \citenamefont {Voigt}, \citenamefont {Lowe}, \citenamefont {Jahangiri},
  \citenamefont {Martin-Delgado}, \citenamefont {Mueller},\ and\ \citenamefont
  {Arrazola}}]{xanadu_batteries}%
  \BibitemOpen
  \bibfield  {author} {\bibinfo {author} {\bibfnamefont {A.}~\bibnamefont
  {Delgado}}, \bibinfo {author} {\bibfnamefont {P.~A.~M.}\ \bibnamefont
  {Casares}}, \bibinfo {author} {\bibfnamefont {R.}~\bibnamefont {dos Reis}},
  \bibinfo {author} {\bibfnamefont {M.~S.}\ \bibnamefont {Zini}}, \bibinfo
  {author} {\bibfnamefont {R.}~\bibnamefont {Campos}}, \bibinfo {author}
  {\bibfnamefont {N.}~\bibnamefont {Cruz-Hern\'andez}}, \bibinfo {author}
  {\bibfnamefont {A.-C.}\ \bibnamefont {Voigt}}, \bibinfo {author}
  {\bibfnamefont {A.}~\bibnamefont {Lowe}}, \bibinfo {author} {\bibfnamefont
  {S.}~\bibnamefont {Jahangiri}}, \bibinfo {author} {\bibfnamefont {M.~A.}\
  \bibnamefont {Martin-Delgado}}, \bibinfo {author} {\bibfnamefont {J.~E.}\
  \bibnamefont {Mueller}}, \ and\ \bibinfo {author} {\bibfnamefont {J.~M.}\
  \bibnamefont {Arrazola}},\ }\href
  {https://link.aps.org/doi/10.1103/PhysRevA.106.032428} {\bibfield  {journal}
  {\bibinfo  {journal} {Phys. Rev. A}\ }\textbf {\bibinfo {volume} {106}},\
  \bibinfo {pages} {032428} (\bibinfo {year} {2022})}\BibitemShut {NoStop}%
\bibitem [{\citenamefont {Rubin}\ \emph {et~al.}(2023)\citenamefont {Rubin},
  \citenamefont {Berry}, \citenamefont {Malone}, \citenamefont {White},
  \citenamefont {Khattar}, \citenamefont {DePrince}, \citenamefont {Sicolo},
  \citenamefont {Kühn}, \citenamefont {Kaicher}, \citenamefont {Lee},\ and\
  \citenamefont {Babbush}}]{bloch_thc}%
  \BibitemOpen
  \bibfield  {author} {\bibinfo {author} {\bibfnamefont {N.~C.}\ \bibnamefont
  {Rubin}}, \bibinfo {author} {\bibfnamefont {D.~W.}\ \bibnamefont {Berry}},
  \bibinfo {author} {\bibfnamefont {F.~D.}\ \bibnamefont {Malone}}, \bibinfo
  {author} {\bibfnamefont {A.~F.}\ \bibnamefont {White}}, \bibinfo {author}
  {\bibfnamefont {T.}~\bibnamefont {Khattar}}, \bibinfo {author} {\bibfnamefont
  {A.~E.}\ \bibnamefont {DePrince}}, \bibinfo {author} {\bibfnamefont
  {S.}~\bibnamefont {Sicolo}}, \bibinfo {author} {\bibfnamefont
  {M.}~\bibnamefont {Kühn}}, \bibinfo {author} {\bibfnamefont
  {M.}~\bibnamefont {Kaicher}}, \bibinfo {author} {\bibfnamefont
  {J.}~\bibnamefont {Lee}}, \ and\ \bibinfo {author} {\bibfnamefont
  {R.}~\bibnamefont {Babbush}},\ }\href {https://arxiv.org/abs/2302.05531}
  {\enquote {\bibinfo {title} {Fault-tolerant quantum simulation of materials
  using bloch orbitals},}\ } (\bibinfo {year} {2023}),\ \Eprint
  {http://arxiv.org/abs/2302.05531} {arXiv:2302.05531 [quant-ph]} \BibitemShut
  {NoStop}%
\bibitem [{\citenamefont {Steudtner}\ \emph {et~al.}(2023)\citenamefont
  {Steudtner}, \citenamefont {Morley-Short}, \citenamefont {Pol}, \citenamefont
  {Sim}, \citenamefont {Cortes}, \citenamefont {Loipersberger}, \citenamefont
  {Parrish}, \citenamefont {Degroote}, \citenamefont {Moll}, \citenamefont
  {Santagati},\ and\ \citenamefont {Streif}}]{mol_obs}%
  \BibitemOpen
  \bibfield  {author} {\bibinfo {author} {\bibfnamefont {M.}~\bibnamefont
  {Steudtner}}, \bibinfo {author} {\bibfnamefont {S.}~\bibnamefont
  {Morley-Short}}, \bibinfo {author} {\bibfnamefont {W.}~\bibnamefont {Pol}},
  \bibinfo {author} {\bibfnamefont {S.}~\bibnamefont {Sim}}, \bibinfo {author}
  {\bibfnamefont {C.~L.}\ \bibnamefont {Cortes}}, \bibinfo {author}
  {\bibfnamefont {M.}~\bibnamefont {Loipersberger}}, \bibinfo {author}
  {\bibfnamefont {R.~M.}\ \bibnamefont {Parrish}}, \bibinfo {author}
  {\bibfnamefont {M.}~\bibnamefont {Degroote}}, \bibinfo {author}
  {\bibfnamefont {N.}~\bibnamefont {Moll}}, \bibinfo {author} {\bibfnamefont
  {R.}~\bibnamefont {Santagati}}, \ and\ \bibinfo {author} {\bibfnamefont
  {M.}~\bibnamefont {Streif}},\ }\href {https://arxiv.org/abs/2303.14118}
  {\enquote {\bibinfo {title} {Fault-tolerant quantum computation of molecular
  observables},}\ } (\bibinfo {year} {2023}),\ \Eprint
  {http://arxiv.org/abs/2303.14118} {arXiv:2303.14118 [quant-ph]} \BibitemShut
  {NoStop}%
\bibitem [{\citenamefont {Kim}\ \emph {et~al.}(2022)\citenamefont {Kim},
  \citenamefont {Liu}, \citenamefont {Pallister}, \citenamefont {Pol},
  \citenamefont {Roberts},\ and\ \citenamefont {Lee}}]{ft_li_battery}%
  \BibitemOpen
  \bibfield  {author} {\bibinfo {author} {\bibfnamefont {I.~H.}\ \bibnamefont
  {Kim}}, \bibinfo {author} {\bibfnamefont {Y.-H.}\ \bibnamefont {Liu}},
  \bibinfo {author} {\bibfnamefont {S.}~\bibnamefont {Pallister}}, \bibinfo
  {author} {\bibfnamefont {W.}~\bibnamefont {Pol}}, \bibinfo {author}
  {\bibfnamefont {S.}~\bibnamefont {Roberts}}, \ and\ \bibinfo {author}
  {\bibfnamefont {E.}~\bibnamefont {Lee}},\ }\href
  {https://link.aps.org/doi/10.1103/PhysRevResearch.4.023019} {\bibfield
  {journal} {\bibinfo  {journal} {Phys. Rev. Res.}\ }\textbf {\bibinfo {volume}
  {4}},\ \bibinfo {pages} {023019} (\bibinfo {year} {2022})}\BibitemShut
  {NoStop}%
\bibitem [{\citenamefont {Goings}\ \emph {et~al.}(2022)\citenamefont {Goings},
  \citenamefont {White}, \citenamefont {Lee}, \citenamefont {Tautermann},
  \citenamefont {Degroote}, \citenamefont {Gidney}, \citenamefont {Shiozaki},
  \citenamefont {Babbush},\ and\ \citenamefont {Rubin}}]{p450}%
  \BibitemOpen
  \bibfield  {author} {\bibinfo {author} {\bibfnamefont {J.~J.}\ \bibnamefont
  {Goings}}, \bibinfo {author} {\bibfnamefont {A.}~\bibnamefont {White}},
  \bibinfo {author} {\bibfnamefont {J.}~\bibnamefont {Lee}}, \bibinfo {author}
  {\bibfnamefont {C.~S.}\ \bibnamefont {Tautermann}}, \bibinfo {author}
  {\bibfnamefont {M.}~\bibnamefont {Degroote}}, \bibinfo {author}
  {\bibfnamefont {C.}~\bibnamefont {Gidney}}, \bibinfo {author} {\bibfnamefont
  {T.}~\bibnamefont {Shiozaki}}, \bibinfo {author} {\bibfnamefont
  {R.}~\bibnamefont {Babbush}}, \ and\ \bibinfo {author} {\bibfnamefont
  {N.~C.}\ \bibnamefont {Rubin}},\ }\href
  {https://doi.org/10.1073%2Fpnas.2203533119} {\bibfield  {journal} {\bibinfo
  {journal} {Proc. Natl. Acad. Sci.}\ }\textbf {\bibinfo {volume} {119}}
  (\bibinfo {year} {2022})}\BibitemShut {NoStop}%
\bibitem [{\citenamefont {Loaiza}\ \emph {et~al.}(2023)\citenamefont {Loaiza},
  \citenamefont {Khah}, \citenamefont {Wiebe},\ and\ \citenamefont
  {Izmaylov}}]{loaiza_lcu}%
  \BibitemOpen
  \bibfield  {author} {\bibinfo {author} {\bibfnamefont {I.}~\bibnamefont
  {Loaiza}}, \bibinfo {author} {\bibfnamefont {A.~M.}\ \bibnamefont {Khah}},
  \bibinfo {author} {\bibfnamefont {N.}~\bibnamefont {Wiebe}}, \ and\ \bibinfo
  {author} {\bibfnamefont {A.~F.}\ \bibnamefont {Izmaylov}},\ }\href
  {https://doi.org/10.1088%2F2058-9565%2Facd577} {\bibfield  {journal}
  {\bibinfo  {journal} {Quantum Sci. Tech.}\ }\textbf {\bibinfo {volume} {8}},\
  \bibinfo {pages} {035019} (\bibinfo {year} {2023})}\BibitemShut {NoStop}%
\bibitem [{\citenamefont {von Burg}\ \emph {et~al.}(2021)\citenamefont {von
  Burg}, \citenamefont {Low}, \citenamefont {Haner}, \citenamefont {Steiger},
  \citenamefont {Reiher}, \citenamefont {Roetteler},\ and\ \citenamefont
  {Troyer}}]{femoco_df}%
  \BibitemOpen
  \bibfield  {author} {\bibinfo {author} {\bibfnamefont {V.}~\bibnamefont {von
  Burg}}, \bibinfo {author} {\bibfnamefont {G.~H.}\ \bibnamefont {Low}},
  \bibinfo {author} {\bibfnamefont {T.}~\bibnamefont {Haner}}, \bibinfo
  {author} {\bibfnamefont {D.}~\bibnamefont {Steiger}}, \bibinfo {author}
  {\bibfnamefont {M.}~\bibnamefont {Reiher}}, \bibinfo {author} {\bibfnamefont
  {M.}~\bibnamefont {Roetteler}}, \ and\ \bibinfo {author} {\bibfnamefont
  {M.}~\bibnamefont {Troyer}},\ }\href
  {https://doi.org/10.1103/PhysRevResearch.3.033055} {\bibfield  {journal}
  {\bibinfo  {journal} {Phys. Rev. Research}\ }\textbf {\bibinfo {volume}
  {3}},\ \bibinfo {pages} {033055} (\bibinfo {year} {2021})}\BibitemShut
  {NoStop}%
\bibitem [{\citenamefont {Campbell}(2021)}]{campbell_hubbard}%
  \BibitemOpen
  \bibfield  {author} {\bibinfo {author} {\bibfnamefont {E.~T.}\ \bibnamefont
  {Campbell}},\ }\href {https://doi.org/10.1088%2F2058-9565%2Fac3110}
  {\bibfield  {journal} {\bibinfo  {journal} {Quantum Sci. Tech.}\ }\textbf
  {\bibinfo {volume} {7}},\ \bibinfo {pages} {015007} (\bibinfo {year}
  {2021})}\BibitemShut {NoStop}%
\bibitem [{\citenamefont {Setia}\ \emph {et~al.}(2020)\citenamefont {Setia},
  \citenamefont {Chen}, \citenamefont {Rice}, \citenamefont {Mezzacapo},
  \citenamefont {Pistoia},\ and\ \citenamefont
  {Whitfield}}]{molecular_point_group_symmetries}%
  \BibitemOpen
  \bibfield  {author} {\bibinfo {author} {\bibfnamefont {K.}~\bibnamefont
  {Setia}}, \bibinfo {author} {\bibfnamefont {R.}~\bibnamefont {Chen}},
  \bibinfo {author} {\bibfnamefont {J.~E.}\ \bibnamefont {Rice}}, \bibinfo
  {author} {\bibfnamefont {A.}~\bibnamefont {Mezzacapo}}, \bibinfo {author}
  {\bibfnamefont {M.}~\bibnamefont {Pistoia}}, \ and\ \bibinfo {author}
  {\bibfnamefont {J.~D.}\ \bibnamefont {Whitfield}},\ }\href
  {https://doi.org/10.1021%2Facs.jctc.0c00113} {\bibfield  {journal} {\bibinfo
  {journal} {J. Chem. Theor. Comput.}\ }\textbf {\bibinfo {volume} {16}},\
  \bibinfo {pages} {6091} (\bibinfo {year} {2020})}\BibitemShut {NoStop}%
\bibitem [{\citenamefont {Peruzzo}\ \emph {et~al.}(2014)\citenamefont
  {Peruzzo}, \citenamefont {McClean}, \citenamefont {Shadbolt}, \citenamefont
  {Yung}, \citenamefont {Zhou}, \citenamefont {Love}, \citenamefont
  {Aspuru-Guzik},\ and\ \citenamefont {O'Brien}}]{vqe}%
  \BibitemOpen
  \bibfield  {author} {\bibinfo {author} {\bibfnamefont {A.}~\bibnamefont
  {Peruzzo}}, \bibinfo {author} {\bibfnamefont {J.}~\bibnamefont {McClean}},
  \bibinfo {author} {\bibfnamefont {P.}~\bibnamefont {Shadbolt}}, \bibinfo
  {author} {\bibfnamefont {M.-H.}\ \bibnamefont {Yung}}, \bibinfo {author}
  {\bibfnamefont {X.-Q.}\ \bibnamefont {Zhou}}, \bibinfo {author}
  {\bibfnamefont {P.~J.}\ \bibnamefont {Love}}, \bibinfo {author}
  {\bibfnamefont {A.}~\bibnamefont {Aspuru-Guzik}}, \ and\ \bibinfo {author}
  {\bibfnamefont {J.~L.}\ \bibnamefont {O'Brien}},\ }\href
  {https://doi.org/10.1038%2Fncomms5213} {\bibfield  {journal} {\bibinfo
  {journal} {Nature Comm.}\ }\textbf {\bibinfo {volume} {5}} (\bibinfo {year}
  {2014})}\BibitemShut {NoStop}%
\bibitem [{\citenamefont {Anand}\ \emph {et~al.}(2022)\citenamefont {Anand},
  \citenamefont {Schleich}, \citenamefont {Alperin-Lea}, \citenamefont
  {Jensen}, \citenamefont {Sim}, \citenamefont {D{\'{\i} }az-Tinoco},
  \citenamefont {Kottmann}, \citenamefont {Degroote}, \citenamefont
  {Izmaylov},\ and\ \citenamefont {Aspuru-Guzik}}]{vqe_rev1}%
  \BibitemOpen
  \bibfield  {author} {\bibinfo {author} {\bibfnamefont {A.}~\bibnamefont
  {Anand}}, \bibinfo {author} {\bibfnamefont {P.}~\bibnamefont {Schleich}},
  \bibinfo {author} {\bibfnamefont {S.}~\bibnamefont {Alperin-Lea}}, \bibinfo
  {author} {\bibfnamefont {P.~W.~K.}\ \bibnamefont {Jensen}}, \bibinfo {author}
  {\bibfnamefont {S.}~\bibnamefont {Sim}}, \bibinfo {author} {\bibfnamefont
  {M.}~\bibnamefont {D{\'{\i} }az-Tinoco}}, \bibinfo {author} {\bibfnamefont
  {J.~S.}\ \bibnamefont {Kottmann}}, \bibinfo {author} {\bibfnamefont
  {M.}~\bibnamefont {Degroote}}, \bibinfo {author} {\bibfnamefont {A.~F.}\
  \bibnamefont {Izmaylov}}, \ and\ \bibinfo {author} {\bibfnamefont
  {A.}~\bibnamefont {Aspuru-Guzik}},\ }\href
  {https://doi.org/10.1039%2Fd1cs00932j} {\bibfield  {journal} {\bibinfo
  {journal} {Chem. Soc. Rev.}\ }\textbf {\bibinfo {volume} {51}},\ \bibinfo
  {pages} {1659} (\bibinfo {year} {2022})}\BibitemShut {NoStop}%
\bibitem [{\citenamefont {Tilly}\ \emph {et~al.}(2022)\citenamefont {Tilly},
  \citenamefont {Chen}, \citenamefont {Cao}, \citenamefont {Picozzi},
  \citenamefont {Setia}, \citenamefont {Li}, \citenamefont {Grant},
  \citenamefont {Wossnig}, \citenamefont {Rungger}, \citenamefont {Booth},\
  and\ \citenamefont {Tennyson}}]{vqe_rev2}%
  \BibitemOpen
  \bibfield  {author} {\bibinfo {author} {\bibfnamefont {J.}~\bibnamefont
  {Tilly}}, \bibinfo {author} {\bibfnamefont {H.}~\bibnamefont {Chen}},
  \bibinfo {author} {\bibfnamefont {S.}~\bibnamefont {Cao}}, \bibinfo {author}
  {\bibfnamefont {D.}~\bibnamefont {Picozzi}}, \bibinfo {author} {\bibfnamefont
  {K.}~\bibnamefont {Setia}}, \bibinfo {author} {\bibfnamefont
  {Y.}~\bibnamefont {Li}}, \bibinfo {author} {\bibfnamefont {E.}~\bibnamefont
  {Grant}}, \bibinfo {author} {\bibfnamefont {L.}~\bibnamefont {Wossnig}},
  \bibinfo {author} {\bibfnamefont {I.}~\bibnamefont {Rungger}}, \bibinfo
  {author} {\bibfnamefont {G.~H.}\ \bibnamefont {Booth}}, \ and\ \bibinfo
  {author} {\bibfnamefont {J.}~\bibnamefont {Tennyson}},\ }\href
  {https://doi.org/10.1016%2Fj.physrep.2022.08.003} {\bibfield  {journal}
  {\bibinfo  {journal} {Phys. Rep.}\ }\textbf {\bibinfo {volume} {986}},\
  \bibinfo {pages} {1} (\bibinfo {year} {2022})}\BibitemShut {NoStop}%
\bibitem [{\citenamefont {Romero}\ \emph {et~al.}(2018)\citenamefont {Romero},
  \citenamefont {Babbush}, \citenamefont {McClean}, \citenamefont {Hempel},
  \citenamefont {Love},\ and\ \citenamefont {Aspuru-Guzik}}]{ucc}%
  \BibitemOpen
  \bibfield  {author} {\bibinfo {author} {\bibfnamefont {J.}~\bibnamefont
  {Romero}}, \bibinfo {author} {\bibfnamefont {R.}~\bibnamefont {Babbush}},
  \bibinfo {author} {\bibfnamefont {J.~R.}\ \bibnamefont {McClean}}, \bibinfo
  {author} {\bibfnamefont {C.}~\bibnamefont {Hempel}}, \bibinfo {author}
  {\bibfnamefont {P.}~\bibnamefont {Love}}, \ and\ \bibinfo {author}
  {\bibfnamefont {A.}~\bibnamefont {Aspuru-Guzik}},\ }\href
  {https://arxiv.org/abs/1701.02691} {\enquote {\bibinfo {title} {Strategies
  for quantum computing molecular energies using the unitary coupled cluster
  ansatz},}\ } (\bibinfo {year} {2018}),\ \Eprint
  {http://arxiv.org/abs/1701.02691} {arXiv:1701.02691 [quant-ph]} \BibitemShut
  {NoStop}%
\bibitem [{\citenamefont {Ryabinkin}\ \emph {et~al.}(2018)\citenamefont
  {Ryabinkin}, \citenamefont {Yen}, \citenamefont {Genin},\ and\ \citenamefont
  {Izmaylov}}]{qcc}%
  \BibitemOpen
  \bibfield  {author} {\bibinfo {author} {\bibfnamefont {I.~G.}\ \bibnamefont
  {Ryabinkin}}, \bibinfo {author} {\bibfnamefont {T.-C.}\ \bibnamefont {Yen}},
  \bibinfo {author} {\bibfnamefont {S.~N.}\ \bibnamefont {Genin}}, \ and\
  \bibinfo {author} {\bibfnamefont {A.~F.}\ \bibnamefont {Izmaylov}},\ }\href
  {https://doi.org/10.1021/acs.jctc.8b00932} {\bibfield  {journal} {\bibinfo
  {journal} {J. Chem. Theor. Comput.}\ }\textbf {\bibinfo {volume} {14}},\
  \bibinfo {pages} {6317} (\bibinfo {year} {2018})}\BibitemShut {NoStop}%
\bibitem [{\citenamefont {Malz}\ \emph {et~al.}(2023)\citenamefont {Malz},
  \citenamefont {Styliaris}, \citenamefont {Wei},\ and\ \citenamefont
  {Cirac}}]{mps_on_qc}%
  \BibitemOpen
  \bibfield  {author} {\bibinfo {author} {\bibfnamefont {D.}~\bibnamefont
  {Malz}}, \bibinfo {author} {\bibfnamefont {G.}~\bibnamefont {Styliaris}},
  \bibinfo {author} {\bibfnamefont {Z.-Y.}\ \bibnamefont {Wei}}, \ and\
  \bibinfo {author} {\bibfnamefont {J.~I.}\ \bibnamefont {Cirac}},\ }\href
  {https://arxiv.org/abs/2307.01696} {\enquote {\bibinfo {title} {Preparation
  of matrix product states with log-depth quantum circuits},}\ } (\bibinfo
  {year} {2023}),\ \Eprint {http://arxiv.org/abs/2307.01696} {arXiv:2307.01696
  [quant-ph]} \BibitemShut {NoStop}%
\bibitem [{\citenamefont {White}(1992)}]{dmrg}%
  \BibitemOpen
  \bibfield  {author} {\bibinfo {author} {\bibfnamefont {S.~R.}\ \bibnamefont
  {White}},\ }\href@noop {} {\bibfield  {journal} {\bibinfo  {journal} {Phys.
  Rev. Lett.}\ }\textbf {\bibinfo {volume} {69}},\ \bibinfo {pages} {2863}
  (\bibinfo {year} {1992})}\BibitemShut {NoStop}%
\bibitem [{\citenamefont {Schollwöck}(2011)}]{dmrg_rev1}%
  \BibitemOpen
  \bibfield  {author} {\bibinfo {author} {\bibfnamefont {U.}~\bibnamefont
  {Schollwöck}},\ }\href {https://doi.org/10.1016%2Fj.aop.2010.09.012}
  {\bibfield  {journal} {\bibinfo  {journal} {Annals Phys.}\ }\textbf {\bibinfo
  {volume} {326}},\ \bibinfo {pages} {96} (\bibinfo {year} {2011})}\BibitemShut
  {NoStop}%
\bibitem [{\citenamefont {Chan}\ and\ \citenamefont
  {Sharma}(2011)}]{dmrg_rev2}%
  \BibitemOpen
  \bibfield  {author} {\bibinfo {author} {\bibfnamefont {G.~K.-L.}\
  \bibnamefont {Chan}}\ and\ \bibinfo {author} {\bibfnamefont {S.}~\bibnamefont
  {Sharma}},\ }\href {https://doi.org/10.1146/annurev-physchem-032210-103338}
  {\bibfield  {journal} {\bibinfo  {journal} {Annual Rev. Phys. Chem.}\
  }\textbf {\bibinfo {volume} {62}},\ \bibinfo {pages} {465} (\bibinfo {year}
  {2011})}\BibitemShut {NoStop}%
\bibitem [{\citenamefont {Silvi}\ \emph {et~al.}(2012)\citenamefont {Silvi},
  \citenamefont {Rossini}, \citenamefont {Fazio}, \citenamefont {Santoro},\
  and\ \citenamefont {Giovannetti}}]{ci_as_mps}%
  \BibitemOpen
  \bibfield  {author} {\bibinfo {author} {\bibfnamefont {P.}~\bibnamefont
  {Silvi}}, \bibinfo {author} {\bibfnamefont {D.}~\bibnamefont {Rossini}},
  \bibinfo {author} {\bibfnamefont {R.}~\bibnamefont {Fazio}}, \bibinfo
  {author} {\bibfnamefont {G.~E.}\ \bibnamefont {Santoro}}, \ and\ \bibinfo
  {author} {\bibfnamefont {V.}~\bibnamefont {Giovannetti}},\ }\href
  {https://doi.org/10.1142%2Fs021797921345029x} {\bibfield  {journal} {\bibinfo
   {journal} {Int. J. Mod. Phys. B}\ }\textbf {\bibinfo {volume} {27}},\
  \bibinfo {pages} {1345029} (\bibinfo {year} {2012})}\BibitemShut {NoStop}%
\bibitem [{\citenamefont {Koridon}\ \emph {et~al.}(2021)\citenamefont
  {Koridon}, \citenamefont {Yalouz}, \citenamefont {Senjean}, \citenamefont
  {Buda}, \citenamefont {O'Brien},\ and\ \citenamefont
  {Visscher}}]{majorana_l1}%
  \BibitemOpen
  \bibfield  {author} {\bibinfo {author} {\bibfnamefont {E.}~\bibnamefont
  {Koridon}}, \bibinfo {author} {\bibfnamefont {S.}~\bibnamefont {Yalouz}},
  \bibinfo {author} {\bibfnamefont {B.}~\bibnamefont {Senjean}}, \bibinfo
  {author} {\bibfnamefont {F.}~\bibnamefont {Buda}}, \bibinfo {author}
  {\bibfnamefont {T.~E.}\ \bibnamefont {O'Brien}}, \ and\ \bibinfo {author}
  {\bibfnamefont {L.}~\bibnamefont {Visscher}},\ }\href
  {https://doi.org/10.1103%2Fphysrevresearch.3.033127} {\bibfield  {journal}
  {\bibinfo  {journal} {Phys. Rev. Research}\ }\textbf {\bibinfo {volume}
  {3}},\ \bibinfo {pages} {033127} (\bibinfo {year} {2021})}\BibitemShut
  {NoStop}%
\bibitem [{\citenamefont {Mogensen}\ and\ \citenamefont
  {Riseth}(2018)}]{optim}%
  \BibitemOpen
  \bibfield  {author} {\bibinfo {author} {\bibfnamefont {P.~K.}\ \bibnamefont
  {Mogensen}}\ and\ \bibinfo {author} {\bibfnamefont {A.~N.}\ \bibnamefont
  {Riseth}},\ }\href {https://doi.org/10.21105/joss.00615} {\bibfield
  {journal} {\bibinfo  {journal} {J. Open Source Softw.}\ }\textbf {\bibinfo
  {volume} {3}},\ \bibinfo {pages} {615} (\bibinfo {year} {2018})}\BibitemShut
  {NoStop}%
\bibitem [{\citenamefont {Foster}\ and\ \citenamefont {Boys}(1960)}]{FB}%
  \BibitemOpen
  \bibfield  {author} {\bibinfo {author} {\bibfnamefont {J.~M.}\ \bibnamefont
  {Foster}}\ and\ \bibinfo {author} {\bibfnamefont {S.~F.}\ \bibnamefont
  {Boys}},\ }\href {https://link.aps.org/doi/10.1103/RevModPhys.32.300}
  {\bibfield  {journal} {\bibinfo  {journal} {Rev. Mod. Phys.}\ }\textbf
  {\bibinfo {volume} {32}},\ \bibinfo {pages} {300} (\bibinfo {year}
  {1960})}\BibitemShut {NoStop}%
\bibitem [{\citenamefont {Crawford}\ \emph {et~al.}(2021)\citenamefont
  {Crawford}, \citenamefont {van Straaten}, \citenamefont {Wang}, \citenamefont
  {Parks}, \citenamefont {Campbell},\ and\ \citenamefont {Brierley}}]{SI}%
  \BibitemOpen
  \bibfield  {author} {\bibinfo {author} {\bibfnamefont {O.}~\bibnamefont
  {Crawford}}, \bibinfo {author} {\bibfnamefont {B.}~\bibnamefont {van
  Straaten}}, \bibinfo {author} {\bibfnamefont {D.}~\bibnamefont {Wang}},
  \bibinfo {author} {\bibfnamefont {T.}~\bibnamefont {Parks}}, \bibinfo
  {author} {\bibfnamefont {E.}~\bibnamefont {Campbell}}, \ and\ \bibinfo
  {author} {\bibfnamefont {S.}~\bibnamefont {Brierley}},\ }\href
  {https://doi.org/10.22331%2Fq-2021-01-20-385} {\bibfield  {journal} {\bibinfo
   {journal} {Quantum}\ }\textbf {\bibinfo {volume} {5}},\ \bibinfo {pages}
  {385} (\bibinfo {year} {2021})}\BibitemShut {NoStop}%
\bibitem [{\citenamefont {Izmaylov}\ \emph {et~al.}(2020)\citenamefont
  {Izmaylov}, \citenamefont {Yen}, \citenamefont {Lang},\ and\ \citenamefont
  {Verteletskyi}}]{anticommuting}%
  \BibitemOpen
  \bibfield  {author} {\bibinfo {author} {\bibfnamefont {A.~F.}\ \bibnamefont
  {Izmaylov}}, \bibinfo {author} {\bibfnamefont {T.-C.}\ \bibnamefont {Yen}},
  \bibinfo {author} {\bibfnamefont {R.~A.}\ \bibnamefont {Lang}}, \ and\
  \bibinfo {author} {\bibfnamefont {V.}~\bibnamefont {Verteletskyi}},\ }\href
  {https://doi.org/10.1021/acs.jctc.9b00791} {\bibfield  {journal} {\bibinfo
  {journal} {J. Chem. Theor. Comput.}\ }\textbf {\bibinfo {volume} {16}},\
  \bibinfo {pages} {190} (\bibinfo {year} {2020})}\BibitemShut {NoStop}%
\bibitem [{\citenamefont {Peng}\ and\ \citenamefont {Kowalski}(2017)}]{df_1}%
  \BibitemOpen
  \bibfield  {author} {\bibinfo {author} {\bibfnamefont {B.}~\bibnamefont
  {Peng}}\ and\ \bibinfo {author} {\bibfnamefont {K.}~\bibnamefont
  {Kowalski}},\ }\href {https://doi.org/10.1021/acs.jctc.7b00605} {\bibfield
  {journal} {\bibinfo  {journal} {J. Chem. Theor. Comput.}\ }\textbf {\bibinfo
  {volume} {13}},\ \bibinfo {pages} {4179} (\bibinfo {year}
  {2017})}\BibitemShut {NoStop}%
\bibitem [{\citenamefont {Motta}\ \emph {et~al.}(2021)\citenamefont {Motta},
  \citenamefont {Ye}, \citenamefont {McClean}, \citenamefont {Li},
  \citenamefont {Minnich}, \citenamefont {Babbush},\ and\ \citenamefont
  {Chan}}]{df_2}%
  \BibitemOpen
  \bibfield  {author} {\bibinfo {author} {\bibfnamefont {M.}~\bibnamefont
  {Motta}}, \bibinfo {author} {\bibfnamefont {E.}~\bibnamefont {Ye}}, \bibinfo
  {author} {\bibfnamefont {J.~R.}\ \bibnamefont {McClean}}, \bibinfo {author}
  {\bibfnamefont {Z.}~\bibnamefont {Li}}, \bibinfo {author} {\bibfnamefont
  {A.~J.}\ \bibnamefont {Minnich}}, \bibinfo {author} {\bibfnamefont
  {R.}~\bibnamefont {Babbush}}, \ and\ \bibinfo {author} {\bibfnamefont
  {G.~K.-L.}\ \bibnamefont {Chan}},\ }\href
  {https://doi.org/10.1038%2Fs41534-021-00416-z} {\bibfield  {journal}
  {\bibinfo  {journal} {npj Quantum Info.}\ }\textbf {\bibinfo {volume} {7}},\
  \bibinfo {pages} {83} (\bibinfo {year} {2021})}\BibitemShut {NoStop}%
\bibitem [{\citenamefont {Motta}\ \emph {et~al.}(2019)\citenamefont {Motta},
  \citenamefont {Shee}, \citenamefont {Zhang},\ and\ \citenamefont
  {Chan}}]{df_3}%
  \BibitemOpen
  \bibfield  {author} {\bibinfo {author} {\bibfnamefont {M.}~\bibnamefont
  {Motta}}, \bibinfo {author} {\bibfnamefont {J.}~\bibnamefont {Shee}},
  \bibinfo {author} {\bibfnamefont {S.}~\bibnamefont {Zhang}}, \ and\ \bibinfo
  {author} {\bibfnamefont {G.~K.-L.}\ \bibnamefont {Chan}},\ }\href
  {https://doi.org/10.1021/acs.jctc.8b00996} {\bibfield  {journal} {\bibinfo
  {journal} {J. Chem. Theor. Comput.}\ }\textbf {\bibinfo {volume} {15}},\
  \bibinfo {pages} {3510} (\bibinfo {year} {2019})}\BibitemShut {NoStop}%
\bibitem [{\citenamefont {Matsuzawa}\ and\ \citenamefont
  {Kurashige}(2020)}]{df_4}%
  \BibitemOpen
  \bibfield  {author} {\bibinfo {author} {\bibfnamefont {Y.}~\bibnamefont
  {Matsuzawa}}\ and\ \bibinfo {author} {\bibfnamefont {Y.}~\bibnamefont
  {Kurashige}},\ }\href {https://doi.org/10.1021/acs.jctc.9b00963} {\bibfield
  {journal} {\bibinfo  {journal} {J. Chem. Theor. Comput.}\ }\textbf {\bibinfo
  {volume} {16}},\ \bibinfo {pages} {944} (\bibinfo {year} {2020})}\BibitemShut
  {NoStop}%
\bibitem [{\citenamefont {Huggins}\ \emph {et~al.}(2021)\citenamefont
  {Huggins}, \citenamefont {McClean}, \citenamefont {Rubin}, \citenamefont
  {Jiang}, \citenamefont {Wiebe}, \citenamefont {Whaley},\ and\ \citenamefont
  {Babbush}}]{df_5}%
  \BibitemOpen
  \bibfield  {author} {\bibinfo {author} {\bibfnamefont {W.~J.}\ \bibnamefont
  {Huggins}}, \bibinfo {author} {\bibfnamefont {J.~R.}\ \bibnamefont
  {McClean}}, \bibinfo {author} {\bibfnamefont {N.~C.}\ \bibnamefont {Rubin}},
  \bibinfo {author} {\bibfnamefont {Z.}~\bibnamefont {Jiang}}, \bibinfo
  {author} {\bibfnamefont {N.}~\bibnamefont {Wiebe}}, \bibinfo {author}
  {\bibfnamefont {K.~B.}\ \bibnamefont {Whaley}}, \ and\ \bibinfo {author}
  {\bibfnamefont {R.}~\bibnamefont {Babbush}},\ }\href
  {https://doi.org/10.1038%2Fs41534-020-00341-7} {\bibfield  {journal}
  {\bibinfo  {journal} {npj Quantum Info.}\ }\textbf {\bibinfo {volume} {7}},\
  \bibinfo {pages} {23} (\bibinfo {year} {2021})}\BibitemShut {NoStop}%
\bibitem [{\citenamefont {Wigner}\ and\ \citenamefont
  {Jordan}(1928)}]{jordan_wigner}%
  \BibitemOpen
  \bibfield  {author} {\bibinfo {author} {\bibfnamefont {E.}~\bibnamefont
  {Wigner}}\ and\ \bibinfo {author} {\bibfnamefont {P.}~\bibnamefont
  {Jordan}},\ }\href
  {https://ui.adsabs.harvard.edu/link_gateway/1928ZPhy...47..631J/doi:10.1007/BF01331938}
  {\bibfield  {journal} {\bibinfo  {journal} {Z. Phys}\ }\textbf {\bibinfo
  {volume} {47}},\ \bibinfo {pages} {631} (\bibinfo {year} {1928})}\BibitemShut
  {NoStop}%
\bibitem [{\citenamefont {Bravyi}\ and\ \citenamefont {Kitaev}(2002)}]{bk1}%
  \BibitemOpen
  \bibfield  {author} {\bibinfo {author} {\bibfnamefont {S.~B.}\ \bibnamefont
  {Bravyi}}\ and\ \bibinfo {author} {\bibfnamefont {A.~Y.}\ \bibnamefont
  {Kitaev}},\ }\href {https://doi.org/10.1006/aphy.2002.6254} {\bibfield
  {journal} {\bibinfo  {journal} {Ann. Phys.}\ }\textbf {\bibinfo {volume}
  {298}},\ \bibinfo {pages} {210} (\bibinfo {year} {2002})}\BibitemShut
  {NoStop}%
\bibitem [{\citenamefont {Seeley}\ \emph {et~al.}(2012)\citenamefont {Seeley},
  \citenamefont {Richard},\ and\ \citenamefont {Love}}]{bk2}%
  \BibitemOpen
  \bibfield  {author} {\bibinfo {author} {\bibfnamefont {J.~T.}\ \bibnamefont
  {Seeley}}, \bibinfo {author} {\bibfnamefont {M.~J.}\ \bibnamefont {Richard}},
  \ and\ \bibinfo {author} {\bibfnamefont {P.~J.}\ \bibnamefont {Love}},\
  }\href {https://doi.org/10.1063/1.4768229} {\bibfield  {journal} {\bibinfo
  {journal} {J. Chem. Phys.}\ }\textbf {\bibinfo {volume} {137}},\ \bibinfo
  {pages} {224109} (\bibinfo {year} {2012})}\BibitemShut {NoStop}%
\bibitem [{\citenamefont {Tranter}\ \emph {et~al.}(2015)\citenamefont
  {Tranter}, \citenamefont {Sofia}, \citenamefont {Seeley}, \citenamefont
  {Kaicher}, \citenamefont {McClean}, \citenamefont {Babbush}, \citenamefont
  {Coveney}, \citenamefont {Mintert}, \citenamefont {Wilhelm},\ and\
  \citenamefont {Love}}]{bk3}%
  \BibitemOpen
  \bibfield  {author} {\bibinfo {author} {\bibfnamefont {A.}~\bibnamefont
  {Tranter}}, \bibinfo {author} {\bibfnamefont {S.}~\bibnamefont {Sofia}},
  \bibinfo {author} {\bibfnamefont {J.}~\bibnamefont {Seeley}}, \bibinfo
  {author} {\bibfnamefont {M.}~\bibnamefont {Kaicher}}, \bibinfo {author}
  {\bibfnamefont {J.}~\bibnamefont {McClean}}, \bibinfo {author} {\bibfnamefont
  {R.}~\bibnamefont {Babbush}}, \bibinfo {author} {\bibfnamefont {P.~V.}\
  \bibnamefont {Coveney}}, \bibinfo {author} {\bibfnamefont {F.}~\bibnamefont
  {Mintert}}, \bibinfo {author} {\bibfnamefont {F.}~\bibnamefont {Wilhelm}}, \
  and\ \bibinfo {author} {\bibfnamefont {P.~J.}\ \bibnamefont {Love}},\ }\href
  {https://doi.org/10.1002/qua.24969} {\bibfield  {journal} {\bibinfo
  {journal} {Int. J. Quantum Chem.}\ }\textbf {\bibinfo {volume} {115}},\
  \bibinfo {pages} {1431} (\bibinfo {year} {2015})}\BibitemShut {NoStop}%
\bibitem [{\citenamefont {Yen}\ and\ \citenamefont {Izmaylov}(2021)}]{CSA}%
  \BibitemOpen
  \bibfield  {author} {\bibinfo {author} {\bibfnamefont {T.-C.}\ \bibnamefont
  {Yen}}\ and\ \bibinfo {author} {\bibfnamefont {A.~F.}\ \bibnamefont
  {Izmaylov}},\ }\href {http://dx.doi.org/10.1103/PRXQuantum.2.040320}
  {\bibfield  {journal} {\bibinfo  {journal} {PRX Quantum}\ }\textbf {\bibinfo
  {volume} {2}},\ \bibinfo {pages} {040320} (\bibinfo {year}
  {2021})}\BibitemShut {NoStop}%
\bibitem [{\citenamefont {Cohn}\ \emph {et~al.}(2021)\citenamefont {Cohn},
  \citenamefont {Motta},\ and\ \citenamefont {Parrish}}]{parrish_csa}%
  \BibitemOpen
  \bibfield  {author} {\bibinfo {author} {\bibfnamefont {J.}~\bibnamefont
  {Cohn}}, \bibinfo {author} {\bibfnamefont {M.}~\bibnamefont {Motta}}, \ and\
  \bibinfo {author} {\bibfnamefont {R.~M.}\ \bibnamefont {Parrish}},\ }\href
  {https://doi.org/10.1103%2Fprxquantum.2.040352} {\bibfield  {journal}
  {\bibinfo  {journal} {{PRX} Quantum}\ }\textbf {\bibinfo {volume} {2}},\
  \bibinfo {pages} {040352} (\bibinfo {year} {2021})}\BibitemShut {NoStop}%
\bibitem [{\citenamefont {Oumarou}\ \emph {et~al.}(2022)\citenamefont
  {Oumarou}, \citenamefont {Scheurer}, \citenamefont {Parrish}, \citenamefont
  {Hohenstein},\ and\ \citenamefont {Gogolin}}]{csa_constrained}%
  \BibitemOpen
  \bibfield  {author} {\bibinfo {author} {\bibfnamefont {O.}~\bibnamefont
  {Oumarou}}, \bibinfo {author} {\bibfnamefont {M.}~\bibnamefont {Scheurer}},
  \bibinfo {author} {\bibfnamefont {R.~M.}\ \bibnamefont {Parrish}}, \bibinfo
  {author} {\bibfnamefont {E.~G.}\ \bibnamefont {Hohenstein}}, \ and\ \bibinfo
  {author} {\bibfnamefont {C.}~\bibnamefont {Gogolin}},\ }\href
  {https://arxiv.org/abs/2212.07957} {\enquote {\bibinfo {title} {Accelerating
  quantum computations of chemistry through regularized compressed double
  factorization},}\ } (\bibinfo {year} {2022}),\ \Eprint
  {http://arxiv.org/abs/2212.07957} {arXiv:2212.07957 [quant-ph]} \BibitemShut
  {NoStop}%
\bibitem [{\citenamefont {Fletcher}(2000)}]{bfgs}%
  \BibitemOpen
  \bibfield  {author} {\bibinfo {author} {\bibfnamefont {R.}~\bibnamefont
  {Fletcher}},\ }\href {https://books.google.ca/books?id=z3m\_EAAAQBAJ} {\emph
  {\bibinfo {title} {Practical Methods of Optimization}}}\ (\bibinfo
  {publisher} {John Wiley \& Sons},\ \bibinfo {address} {West Sussex,
  England},\ \bibinfo {year} {2000})\BibitemShut {NoStop}%
\bibitem [{\citenamefont {Dunning}\ \emph {et~al.}(2017)\citenamefont
  {Dunning}, \citenamefont {Huchette},\ and\ \citenamefont {Lubin}}]{jump}%
  \BibitemOpen
  \bibfield  {author} {\bibinfo {author} {\bibfnamefont {I.}~\bibnamefont
  {Dunning}}, \bibinfo {author} {\bibfnamefont {J.}~\bibnamefont {Huchette}}, \
  and\ \bibinfo {author} {\bibfnamefont {M.}~\bibnamefont {Lubin}},\ }\href
  {\doibase 10.1137/15M1020575} {\bibfield  {journal} {\bibinfo  {journal}
  {SIAM Review}\ }\textbf {\bibinfo {volume} {59}},\ \bibinfo {pages} {295}
  (\bibinfo {year} {2017})}\BibitemShut {NoStop}%
\bibitem [{\citenamefont {Huangfu}\ and\ \citenamefont {Hall}(2018)}]{highs}%
  \BibitemOpen
  \bibfield  {author} {\bibinfo {author} {\bibfnamefont {Q.}~\bibnamefont
  {Huangfu}}\ and\ \bibinfo {author} {\bibfnamefont {J.}~\bibnamefont {Hall}},\
  }\href {\doibase 10.1007/s12532-017-0130-5} {\bibfield  {journal} {\bibinfo
  {journal} {Math. Prog. Comput.}\ }\textbf {\bibinfo {volume} {10}},\ \bibinfo
  {pages} {119–142} (\bibinfo {year} {2018})}\BibitemShut {NoStop}%
\bibitem [{\citenamefont {Sun}(2015)}]{pyscf1}%
  \BibitemOpen
  \bibfield  {author} {\bibinfo {author} {\bibfnamefont {Q.}~\bibnamefont
  {Sun}},\ }\href {https://onlinelibrary.wiley.com/doi/abs/10.1002/jcc.23981}
  {\bibfield  {journal} {\bibinfo  {journal} {J. Comput. Chem.}\ }\textbf
  {\bibinfo {volume} {36}},\ \bibinfo {pages} {1664} (\bibinfo {year}
  {2015})}\BibitemShut {NoStop}%
\bibitem [{\citenamefont {Sun}\ \emph {et~al.}(2018)\citenamefont {Sun},
  \citenamefont {Berkelbach}, \citenamefont {Blunt}, \citenamefont {Booth},
  \citenamefont {Guo}, \citenamefont {Li}, \citenamefont {Liu}, \citenamefont
  {McClain}, \citenamefont {Sayfutyarova}, \citenamefont {Sharma},
  \citenamefont {Wouters},\ and\ \citenamefont {Chan}}]{pyscf2}%
  \BibitemOpen
  \bibfield  {author} {\bibinfo {author} {\bibfnamefont {Q.}~\bibnamefont
  {Sun}}, \bibinfo {author} {\bibfnamefont {T.~C.}\ \bibnamefont {Berkelbach}},
  \bibinfo {author} {\bibfnamefont {N.~S.}\ \bibnamefont {Blunt}}, \bibinfo
  {author} {\bibfnamefont {G.~H.}\ \bibnamefont {Booth}}, \bibinfo {author}
  {\bibfnamefont {S.}~\bibnamefont {Guo}}, \bibinfo {author} {\bibfnamefont
  {Z.}~\bibnamefont {Li}}, \bibinfo {author} {\bibfnamefont {J.}~\bibnamefont
  {Liu}}, \bibinfo {author} {\bibfnamefont {J.~D.}\ \bibnamefont {McClain}},
  \bibinfo {author} {\bibfnamefont {E.~R.}\ \bibnamefont {Sayfutyarova}},
  \bibinfo {author} {\bibfnamefont {S.}~\bibnamefont {Sharma}}, \bibinfo
  {author} {\bibfnamefont {S.}~\bibnamefont {Wouters}}, \ and\ \bibinfo
  {author} {\bibfnamefont {G.~K.-L.}\ \bibnamefont {Chan}},\ }\href
  {https://wires.onlinelibrary.wiley.com/doi/abs/10.1002/wcms.1340} {\bibfield
  {journal} {\bibinfo  {journal} {WIREs Comput. Mol. Sci.}\ }\textbf {\bibinfo
  {volume} {8}},\ \bibinfo {pages} {e1340} (\bibinfo {year}
  {2018})}\BibitemShut {NoStop}%
\bibitem [{\citenamefont {Sun}\ \emph {et~al.}(2020)\citenamefont {Sun},
  \citenamefont {Zhang}, \citenamefont {Banerjee}, \citenamefont {Bao},
  \citenamefont {Barbry}, \citenamefont {Blunt}, \citenamefont {Bogdanov},
  \citenamefont {Booth}, \citenamefont {Chen}, \citenamefont {Cui},
  \citenamefont {Eriksen}, \citenamefont {Gao}, \citenamefont {Guo},
  \citenamefont {Hermann}, \citenamefont {Hermes}, \citenamefont {Koh},
  \citenamefont {Koval}, \citenamefont {Lehtola}, \citenamefont {Li},
  \citenamefont {Liu}, \citenamefont {Mardirossian}, \citenamefont {McClain},
  \citenamefont {Motta}, \citenamefont {Mussard}, \citenamefont {Pham},
  \citenamefont {Pulkin}, \citenamefont {Purwanto}, \citenamefont {Robinson},
  \citenamefont {Ronca}, \citenamefont {Sayfutyarova}, \citenamefont
  {Scheurer}, \citenamefont {Schurkus}, \citenamefont {Smith}, \citenamefont
  {Sun}, \citenamefont {Sun}, \citenamefont {Upadhyay}, \citenamefont {Wagner},
  \citenamefont {Wang}, \citenamefont {White}, \citenamefont {Whitfield},
  \citenamefont {Williamson}, \citenamefont {Wouters}, \citenamefont {Yang},
  \citenamefont {Yu}, \citenamefont {Zhu}, \citenamefont {Berkelbach},
  \citenamefont {Sharma}, \citenamefont {Sokolov},\ and\ \citenamefont
  {Chan}}]{pyscf3}%
  \BibitemOpen
  \bibfield  {author} {\bibinfo {author} {\bibfnamefont {Q.}~\bibnamefont
  {Sun}}, \bibinfo {author} {\bibfnamefont {X.}~\bibnamefont {Zhang}}, \bibinfo
  {author} {\bibfnamefont {S.}~\bibnamefont {Banerjee}}, \bibinfo {author}
  {\bibfnamefont {P.}~\bibnamefont {Bao}}, \bibinfo {author} {\bibfnamefont
  {M.}~\bibnamefont {Barbry}}, \bibinfo {author} {\bibfnamefont {N.~S.}\
  \bibnamefont {Blunt}}, \bibinfo {author} {\bibfnamefont {N.~A.}\ \bibnamefont
  {Bogdanov}}, \bibinfo {author} {\bibfnamefont {G.~H.}\ \bibnamefont {Booth}},
  \bibinfo {author} {\bibfnamefont {J.}~\bibnamefont {Chen}}, \bibinfo {author}
  {\bibfnamefont {Z.-H.}\ \bibnamefont {Cui}}, \bibinfo {author} {\bibfnamefont
  {J.~J.}\ \bibnamefont {Eriksen}}, \bibinfo {author} {\bibfnamefont
  {Y.}~\bibnamefont {Gao}}, \bibinfo {author} {\bibfnamefont {S.}~\bibnamefont
  {Guo}}, \bibinfo {author} {\bibfnamefont {J.}~\bibnamefont {Hermann}},
  \bibinfo {author} {\bibfnamefont {M.~R.}\ \bibnamefont {Hermes}}, \bibinfo
  {author} {\bibfnamefont {K.}~\bibnamefont {Koh}}, \bibinfo {author}
  {\bibfnamefont {P.}~\bibnamefont {Koval}}, \bibinfo {author} {\bibfnamefont
  {S.}~\bibnamefont {Lehtola}}, \bibinfo {author} {\bibfnamefont
  {Z.}~\bibnamefont {Li}}, \bibinfo {author} {\bibfnamefont {J.}~\bibnamefont
  {Liu}}, \bibinfo {author} {\bibfnamefont {N.}~\bibnamefont {Mardirossian}},
  \bibinfo {author} {\bibfnamefont {J.~D.}\ \bibnamefont {McClain}}, \bibinfo
  {author} {\bibfnamefont {M.}~\bibnamefont {Motta}}, \bibinfo {author}
  {\bibfnamefont {B.}~\bibnamefont {Mussard}}, \bibinfo {author} {\bibfnamefont
  {H.~Q.}\ \bibnamefont {Pham}}, \bibinfo {author} {\bibfnamefont
  {A.}~\bibnamefont {Pulkin}}, \bibinfo {author} {\bibfnamefont
  {W.}~\bibnamefont {Purwanto}}, \bibinfo {author} {\bibfnamefont {P.~J.}\
  \bibnamefont {Robinson}}, \bibinfo {author} {\bibfnamefont {E.}~\bibnamefont
  {Ronca}}, \bibinfo {author} {\bibfnamefont {E.~R.}\ \bibnamefont
  {Sayfutyarova}}, \bibinfo {author} {\bibfnamefont {M.}~\bibnamefont
  {Scheurer}}, \bibinfo {author} {\bibfnamefont {H.~F.}\ \bibnamefont
  {Schurkus}}, \bibinfo {author} {\bibfnamefont {J.~E.~T.}\ \bibnamefont
  {Smith}}, \bibinfo {author} {\bibfnamefont {C.}~\bibnamefont {Sun}}, \bibinfo
  {author} {\bibfnamefont {S.-N.}\ \bibnamefont {Sun}}, \bibinfo {author}
  {\bibfnamefont {S.}~\bibnamefont {Upadhyay}}, \bibinfo {author}
  {\bibfnamefont {L.~K.}\ \bibnamefont {Wagner}}, \bibinfo {author}
  {\bibfnamefont {X.}~\bibnamefont {Wang}}, \bibinfo {author} {\bibfnamefont
  {A.}~\bibnamefont {White}}, \bibinfo {author} {\bibfnamefont {J.~D.}\
  \bibnamefont {Whitfield}}, \bibinfo {author} {\bibfnamefont {M.~J.}\
  \bibnamefont {Williamson}}, \bibinfo {author} {\bibfnamefont
  {S.}~\bibnamefont {Wouters}}, \bibinfo {author} {\bibfnamefont
  {J.}~\bibnamefont {Yang}}, \bibinfo {author} {\bibfnamefont {J.~M.}\
  \bibnamefont {Yu}}, \bibinfo {author} {\bibfnamefont {T.}~\bibnamefont
  {Zhu}}, \bibinfo {author} {\bibfnamefont {T.~C.}\ \bibnamefont {Berkelbach}},
  \bibinfo {author} {\bibfnamefont {S.}~\bibnamefont {Sharma}}, \bibinfo
  {author} {\bibfnamefont {A.~Y.}\ \bibnamefont {Sokolov}}, \ and\ \bibinfo
  {author} {\bibfnamefont {G.~K.-L.}\ \bibnamefont {Chan}},\ }\href
  {https://doi.org/10.1063/5.0006074} {\bibfield  {journal} {\bibinfo
  {journal} {J. Chem. Phys.}\ }\textbf {\bibinfo {volume} {153}},\ \bibinfo
  {pages} {024109} (\bibinfo {year} {2020})}\BibitemShut {NoStop}%
\bibitem [{\citenamefont {McClean}\ \emph {et~al.}(2020)\citenamefont
  {McClean}, \citenamefont {Rubin}, \citenamefont {Sung}, \citenamefont
  {Kivlichan}, \citenamefont {Bonet-Monroig}, \citenamefont {Cao},
  \citenamefont {Dai}, \citenamefont {Fried}, \citenamefont {Gidney},
  \citenamefont {Gimby}, \citenamefont {Gokhale}, \citenamefont {Häner},
  \citenamefont {Hardikar}, \citenamefont {Havl{\'{\i}}{\v{c}}ek},
  \citenamefont {Higgott}, \citenamefont {Huang}, \citenamefont {Izaac},
  \citenamefont {Jiang}, \citenamefont {Liu}, \citenamefont {McArdle},
  \citenamefont {Neeley}, \citenamefont {O'Brien}, \citenamefont {O'Gorman},
  \citenamefont {Ozfidan}, \citenamefont {Radin}, \citenamefont {Romero},
  \citenamefont {Sawaya}, \citenamefont {Senjean}, \citenamefont {Setia},
  \citenamefont {Sim}, \citenamefont {Steiger}, \citenamefont {Steudtner},
  \citenamefont {Sun}, \citenamefont {Sun}, \citenamefont {Wang}, \citenamefont
  {Zhang},\ and\ \citenamefont {Babbush}}]{openfermion}%
  \BibitemOpen
  \bibfield  {author} {\bibinfo {author} {\bibfnamefont {J.~R.}\ \bibnamefont
  {McClean}}, \bibinfo {author} {\bibfnamefont {N.~C.}\ \bibnamefont {Rubin}},
  \bibinfo {author} {\bibfnamefont {K.~J.}\ \bibnamefont {Sung}}, \bibinfo
  {author} {\bibfnamefont {I.~D.}\ \bibnamefont {Kivlichan}}, \bibinfo {author}
  {\bibfnamefont {X.}~\bibnamefont {Bonet-Monroig}}, \bibinfo {author}
  {\bibfnamefont {Y.}~\bibnamefont {Cao}}, \bibinfo {author} {\bibfnamefont
  {C.}~\bibnamefont {Dai}}, \bibinfo {author} {\bibfnamefont {E.~S.}\
  \bibnamefont {Fried}}, \bibinfo {author} {\bibfnamefont {C.}~\bibnamefont
  {Gidney}}, \bibinfo {author} {\bibfnamefont {B.}~\bibnamefont {Gimby}},
  \bibinfo {author} {\bibfnamefont {P.}~\bibnamefont {Gokhale}}, \bibinfo
  {author} {\bibfnamefont {T.}~\bibnamefont {Häner}}, \bibinfo {author}
  {\bibfnamefont {T.}~\bibnamefont {Hardikar}}, \bibinfo {author}
  {\bibfnamefont {V.}~\bibnamefont {Havl{\'{\i}}{\v{c}}ek}}, \bibinfo {author}
  {\bibfnamefont {O.}~\bibnamefont {Higgott}}, \bibinfo {author} {\bibfnamefont
  {C.}~\bibnamefont {Huang}}, \bibinfo {author} {\bibfnamefont
  {J.}~\bibnamefont {Izaac}}, \bibinfo {author} {\bibfnamefont
  {Z.}~\bibnamefont {Jiang}}, \bibinfo {author} {\bibfnamefont
  {X.}~\bibnamefont {Liu}}, \bibinfo {author} {\bibfnamefont {S.}~\bibnamefont
  {McArdle}}, \bibinfo {author} {\bibfnamefont {M.}~\bibnamefont {Neeley}},
  \bibinfo {author} {\bibfnamefont {T.}~\bibnamefont {O'Brien}}, \bibinfo
  {author} {\bibfnamefont {B.}~\bibnamefont {O'Gorman}}, \bibinfo {author}
  {\bibfnamefont {I.}~\bibnamefont {Ozfidan}}, \bibinfo {author} {\bibfnamefont
  {M.~D.}\ \bibnamefont {Radin}}, \bibinfo {author} {\bibfnamefont
  {J.}~\bibnamefont {Romero}}, \bibinfo {author} {\bibfnamefont {N.~P.~D.}\
  \bibnamefont {Sawaya}}, \bibinfo {author} {\bibfnamefont {B.}~\bibnamefont
  {Senjean}}, \bibinfo {author} {\bibfnamefont {K.}~\bibnamefont {Setia}},
  \bibinfo {author} {\bibfnamefont {S.}~\bibnamefont {Sim}}, \bibinfo {author}
  {\bibfnamefont {D.~S.}\ \bibnamefont {Steiger}}, \bibinfo {author}
  {\bibfnamefont {M.}~\bibnamefont {Steudtner}}, \bibinfo {author}
  {\bibfnamefont {Q.}~\bibnamefont {Sun}}, \bibinfo {author} {\bibfnamefont
  {W.}~\bibnamefont {Sun}}, \bibinfo {author} {\bibfnamefont {D.}~\bibnamefont
  {Wang}}, \bibinfo {author} {\bibfnamefont {F.}~\bibnamefont {Zhang}}, \ and\
  \bibinfo {author} {\bibfnamefont {R.}~\bibnamefont {Babbush}},\ }\href
  {https://doi.org/10.1088/2058-9565/ab8ebc} {\bibfield  {journal} {\bibinfo
  {journal} {Quantum Sci. Tech.}\ }\textbf {\bibinfo {volume} {5}},\ \bibinfo
  {pages} {034014} (\bibinfo {year} {2020})}\BibitemShut {NoStop}%
\bibitem [{\citenamefont {Hehre}\ \emph {et~al.}(1969)\citenamefont {Hehre},
  \citenamefont {Stewart},\ and\ \citenamefont {Pople}}]{sto3g}%
  \BibitemOpen
  \bibfield  {author} {\bibinfo {author} {\bibfnamefont {W.~J.}\ \bibnamefont
  {Hehre}}, \bibinfo {author} {\bibfnamefont {R.~F.}\ \bibnamefont {Stewart}},
  \ and\ \bibinfo {author} {\bibfnamefont {J.~A.}\ \bibnamefont {Pople}},\
  }\href {https://doi.org/10.1063/1.1672392} {\bibfield  {journal} {\bibinfo
  {journal} {J. Chem. Phys.}\ }\textbf {\bibinfo {volume} {51}},\ \bibinfo
  {pages} {2657} (\bibinfo {year} {1969})}\BibitemShut {NoStop}%
\bibitem [{\citenamefont {Szabo}\ and\ \citenamefont {Ostlund}(1996)}]{szabo}%
  \BibitemOpen
  \bibfield  {author} {\bibinfo {author} {\bibfnamefont {A.}~\bibnamefont
  {Szabo}}\ and\ \bibinfo {author} {\bibfnamefont {N.}~\bibnamefont
  {Ostlund}},\ }\href {https://books.google.ca/books?id=k-DcCgAAQBAJ} {\emph
  {\bibinfo {title} {Modern Quantum Chemistry: Introduction to Advanced
  Electronic Structure Theory}}},\ Dover Books on Chemistry\ (\bibinfo
  {publisher} {Dover Publications},\ \bibinfo {address} {Mineola, New York,
  USA},\ \bibinfo {year} {1996})\BibitemShut {NoStop}%
\end{thebibliography}%
